\documentclass[article, shortnames]{jss}
\usepackage{thumbpdf,lmodern} 
\graphicspath{{Figures/}}  

\usepackage{pdfpages}
\usepackage{float}
\usepackage[toc,page]{appendix}
\usepackage{subcaption}
\usepackage{grffile}
\usepackage{amsmath, amssymb}
\usepackage{amsthm}
\usepackage{caption}
\usepackage{accents}
\usepackage[cal=euler,
calscaled=1.0,
bb=ams,
frak=mma,
frakscaled=.97,
scr=rsfs]{mathalfa}
\newcommand\thickbar[1]{\accentset{\rule{.4em}{.8pt}}{#1}}
\newcommand\widebar[1]{\accentset{\rule{.8em}{.8pt}}{#1}}
\newcommand\thickubar[1]{\underaccent{\rule{.45em}{.8pt}}{#1}}
\usepackage[plain]{algorithm} 
\usepackage[noend]{algpseudocode}

\newcommand{\indep}{\,\rotatebox[origin=c]{90}{$\models$}\,}
\newcommand{\+}[1]{\ensuremath{\mathbf{#1}}}

\algnewcommand\algorithmicinput{\textbf{INPUT:}}
\algnewcommand\INPUT{\item[\algorithmicinput]}
\algnewcommand\algorithmicoutput{\textbf{OUTPUT:}}
\algnewcommand\OUTPUT{\item[\algorithmicoutput]}

\theoremstyle{plain}
\theoremstyle{definition}
\newtheorem{definition}{Definition}

\author{Santtu Tikka\\University of Jyvaskyla
   \And Juha Karvanen\\University of Jyvaskyla}
\Plainauthor{Santtu Tikka, Juha Karvanen}

\title{Identifying Causal Effects with the \proglang{R} Package \pkg{causaleffect}}
\Plaintitle{Identifying Causal Effects with the R Package causaleffect}

\Abstract{
Do-calculus is concerned with estimating the interventional distribution of an action from the
observed joint probability distribution of the variables in a given causal structure. All identifiable
causal effects can be derived using the rules of do-calculus, but the rules themselves do not give any
direct indication whether the effect in question is identifiable or not. \cite{shpitser06} constructed
an algorithm for identifying joint interventional distributions in causal models,
which contain unobserved variables and induce directed acyclic graphs. This algorithm can be seen
as a repeated application of the rules of do-calculus and known properties of probabilities, and it
ultimately either derives an expression for the causal distribution, or fails to identify the effect, in
which case the effect is non-identifiable. In this paper, the \proglang{R} package \pkg{causaleffect} is
presented, which provides an implementation of this algorithm. Functionality of \pkg{causaleffect} is
also demonstrated through examples.
}

\Keywords{DAG, do-calculus, causality, causal model, identifiability, graph, C-component, hedge, d-separation}

\Volume{76}
\Issue{12}
\Month{February}
\Year{2017}
\Submitdate{2015-05-06}
\Acceptdate{2015-11-25}
\DOI{10.18637/jss.v076.i12}

\Address{
  Santtu Tikka\\
  Department of Mathematics and Statistics\\
  Faculty of Mathematics and Science\\
  University of Jyvaskyla\\
  P.O.Box 35, FI-40014, Finland\\
  E-mail: \email{santtu.tikka@jyu.fi}\\
}

\begin{document}

\section{Introduction} \label{Sect:intro}
When discussing causality, one often means the relationships between events, where a set of events directly or indirectly causes another set of events. The aim of causal inference is to draw conclusions from these relationships by using available data and prior knowledge. Causal inference can also be applied when determining the effects of actions on some variables of interest. These types of actions are often called interventions and the results of the interventions are referred to as causal effects.

The causal inference can be divided into three sub-areas: discovering the causal model from the data, identifying the causal effect when the causal structure is known and estimating an identifiable causal effect from the data. Our contribution belongs to the second category, identification of causal effects. As a starting point, we assume that the causal relationships between the variables are known in a non-parametric form and formally presented as a probabilistic causal model \citep{pearl95}. Part of the variables may be latent. The causal structure, i.e., the non-parametric causal relationships, can be described using a directed acyclic graph (DAG). A causal effect is called identifiable if it can be uniquely determined from the causal structure on basis of the observations only.

Do-calculus \citep{pearl95} consist of a set of inference rules, which can be used to express the interventional probability distribution using only observational distributions. The rules of do-calculus do not themselves indicate the order in which they should be applied. This problem is solved in the algorithm developed by \citet{tian03} and \citet{shpitser06}. The algorithm is proved to determine the interventional distribution of an identifiable causal effect. When faced with an unidentifiable effect, the algorithm provides a problematic graph structure called a hedge, which can be thought of as the cause of unidentifiability.

Other \proglang{R} packages for causal inference are summarized in Table~\ref{tab:Rpackages}. It can be seen that in addition to \pkg{causaleffect}, only \pkg{pcalg} \citep{pcalg} supports the identification of causal effects. \pkg{pcalg} supports the generalized back-door criterion but does not support the front-door criterion. Thus, according to our knowledge, \pkg{causaleffect} is the only \proglang{R} package that implements a complete algorithm for the identification of causal effects.

An algorithm equivalent to the one developed by \citep{shpitser06} has been
implemented earlier by Lexin Liu in the \pkg{CIBN} software using
\pkg{JavaBayes}, which is a graphical software interface written in
\proglang{Java} by Fabio Gagliardi Cozman. In addition to causal effect
identification \pkg{CIBN} also provides tools for creating and editing
graphical models. \pkg{CIBN} is freely available from
\url{http://web.cs.iastate.edu/~jtian/Software/CIBN.htm}. \pkg{DAGitty}
\citep{dagitty} provides another free interface for causal inference and causal
modeling. One of the main features of \pkg{DAGitty} is finding sufficient
adjustment sets for the minimization of bias in causal effect estimation.
\pkg{DAGitty} can also be used to determine instrumental variables, which is a
feature currently not provided by \pkg{causaleffect}. However, \pkg{DAGitty}
does not provide a complete criterion for identifiability.

Familiarity of Pearl's causal model, do-calculus and basic graph theory is assumed throughout the paper. These concepts are briefly reviewed in Appendix~\ref{App:AppendixA}. A more detailed description can be found in \citep{pearl09} and \citep{koller09}. Notation similar to that of \citep{shpitser06} is also utilized repeatedly in this paper. Capital letters denote variables and small letters denote their values. Bold letters denote sets which are formed of the previous two. The abbreviations $Pa(\+ Y)_G, An(\+ Y)_G, $ and $De(\+ Y)_G$ denote the set of observable parents, ancestors and descendants of the node set $\+ Y$ while also containing $\+ Y$ itself. It should also be noted that the shorthand notation of bidirected edges is used to represent the direct effects of an unobserved confounding variable on the two variables at the endpoints of the bidirected edge.

\begin{table}[p!]
  \begin{tabular}{l p{10.2cm}}
  \hline

  \multicolumn{2}{l}{\it Packages for specific applications} \\
  \hline
  \pkg{ASPBay} & Bayesian inference on causal genetic variants using affected sib-pairs data \citep{ASPBay} \\
  \pkg{cin}    & Causal inference for neuroscience \citep{cin}\\
  \pkg{mwa}    & Causal inference in spatiotemporal event data \citep{mwa}\\
  \pkg{qtlnet} & Causal inference of QTL networks \citep{qtlnet}\\
  \hline

  \multicolumn{2}{l}{\it Packages for estimation of causal effects from data} \\
  \hline
  \pkg{CausalGAM}  &     Estimation of causal effects with generalized additive models \citep{CausalGAM} \\
  \pkg{InvariantCausalPrediction}  &     Invariant causal prediction \citep{InvariantCausalPrediction} \\
  \pkg{iWeigReg}  &      Improved methods for causal inference and missing data problems \citep{iWeigReg} \\
  \pkg{pcalg}  & Methods for graphical models and causal inference \\
  \pkg{SVMMatch}  &      Causal effect estimation and diagnostics with support vector machines \citep{SVMMatch} \\
  \pkg{wfe}  &   Weighted linear fixed effects regression models for causal inference \citep{wfe} \\
  \hline

  \multicolumn{2}{l}{\it Packages for sensitivity analysis and other specific problems in causal inference} \\
  \hline
  \pkg{causalsens}  &      Selection bias approach to sensitivity analysis for causal effects \citep{causalsens} \\
  \pkg{cit}  &     Causal inference test \citep{cit} \\
  \pkg{ImpactIV}  &        Identifying causal effect for multi-component intervention using instrumental variable method \citep{ImpactIV}\\
  \pkg{inferference}  &    Methods for causal inference with interference \citep{inferference} \\
  \pkg{MatchingFrontier}  &        Computation of the balance -- sample size frontier in matching methods for causal inference \citep{MatchingFrontier}\\
  \pkg{mediation}  &       Causal mediation analysis \citep{mediation} \\
  \pkg{qualCI}  & Causal inference with qualitative and ordinal information on outcomes\citep{qualCI} \\
  \pkg{SimpleTable}  &     Bayesian inference and sensitivity analysis for causal effects from 2 $\times$ 2 and 2 $\times$ 2 $\times K$ tables in the presence of unmeasured confounding \citep{SimpleTable} \\
  \pkg{treatSens} &       Sensitivity analysis for causal inference \citep{treatSens} \\
  \hline

  \multicolumn{2}{l}{\it Packages for causal discovery} \\
  \hline
  \pkg{CAM}  &     Causal additive model (CAM) \citep{cam}\\
  \pkg{D2C}  &     Predicting causal direction from dependency features \citep{D2C} \\
  \pkg{pcalg}  &   Methods for graphical models and causal inference \\
  \hline

  \multicolumn{2}{l}{\it Packages for identification of causal effects} \\
  \hline
  \pkg{causaleffect} &  Deriving expressions of joint interventional distributions in causal models \\
  \pkg{pcalg} &   Methods for graphical models and causal inference \\
  \hline
  \end{tabular}
  \caption{\proglang{R} packages for causal inference.}
  \label{tab:Rpackages}
\end{table}

A motivating example is presented in Section~\ref{Sect:docalc}. The identification algorithm is presented in Section~\ref{Sect:algo} and the details of its \proglang{R} implementation are described in Section~\ref{Sect:implementinR}. Section~\ref{Sect:packagece} showcases the usage of \pkg{causaleffect} in \proglang{R} with some simple examples, and describes some curious special cases arising from the nature of the algorithm itself. Section~\ref{Sect:discussion} concludes this paper by providing some examples of similar algorithms, where the work of this paper could be applicable.

\section[Example on do-calculus]{Example on do-calculus} \label{Sect:docalc}

\begin{figure}[t!]
    \centering
    \vspace*{-0.5cm}
    \includegraphics[width=0.27\textwidth]{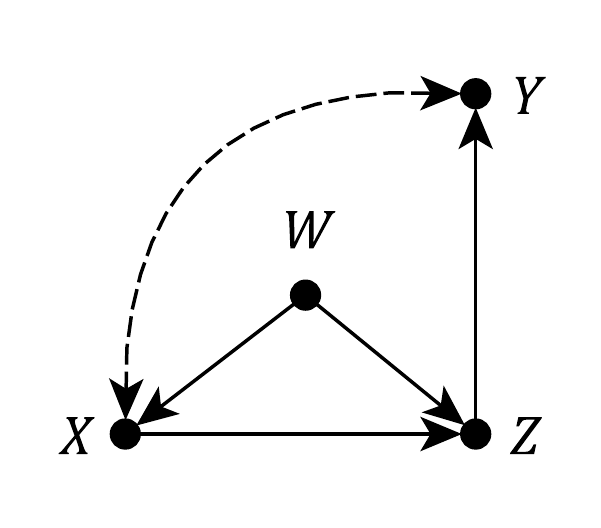}
    \caption{Graph $G$ for the illustrative example.}
    \label{kuva:grfGm}
\end{figure}

Consider identification of causal effect $P_x(y)$ in the graph $G$ of Figure~\ref{kuva:grfGm}. We show how this causal effect can be identified by applying do-calculus \citep{pearl09} manually. Later the same example is reconsidered using the identification algorithm.

First, the rules of do-calculus are shortly reviewed. The purpose of do-calculus is to represent the interventional distribution $P_{\+ x}(\+y)$ by using only observational probabilities. A causal effect is identifiable, if such an expression can be found by applying the rules of do-calculus repeatedly. This result follow directly from the definition of identifiability due to the fact that all observational distributions are assumed identical for the causal models that induce $G$.

Let $\+X, \+Y$ and $\+ Z$ be pairwise disjoint sets of nodes in the graph $G$ induced by a causal model $M$. Here $G_{\thickbar {\+ X},\thickubar {\+ Z}}$ means the graph that is obtained from $G$ by removing all incoming edges of $\+ X$ and all outgoing edges of $\+ Z$. Let $P$ be the joint distribution of all observed and unobserved variables of $M$. Now, the following three rules hold \citep{pearl95}:
\begin{enumerate}
\item{Insertion and deletion of observations: 
$$P_{\+x}(\+ y|\+ z, \+ w) = P_{\+x}(\+ y| \+ w), \text{ if } (\+ Y \indep \+Z|\+X, \+ W)_{G_{\thickbar {\+ X}}}.$$}
\item{Exchanging actions and observations:
$$P_{\+x,\+ z}(\+ y|\+ w) = P_{\+x}(\+ y|\+ z, \+ w), \text{ if } (\+ Y \indep \+Z|\+X, \+ W)_{G_{\thickbar {\+ X},\thickubar {\+ Z}}}.$$ }
\item{Insertion and deletion of actions:
$$P_{\+x,\+ z}(\+ y|\+ w) = P_{\+x}(\+ y|\+ w), \text{ if } (\+ Y \indep \+Z|\+X, \+ W)_{G_{\thickbar {\+ X},\widebar {Z(\+ W)}}},$$ 
where $ Z(\+W) = \+Z \setminus An(\+ W)_{G_{\thickbar {\+ X}}}.$ }
\end{enumerate}
The rules of do-calculus can be shown to be true by using d-separation and the definition of the $do(\cdot)$-operator. Pearl presented proofs for these three rules \citep{pearl95}. Do-calculus has also been shown to be complete, meaning that the expressions of all identifiable causal effects can be derived by using the three rules \citep{shpitser06,Pearlsiscomplete}.

To identify $P_x(y)$ in the causal model of Figure~\ref{kuva:grfGm}, we begin with the factorization
\begin{equation} \label{eq:big1}
P_x(y) = \sum_{w,z}P_x(y|w,z)P_x(z|w)P_x(w).
\end{equation}

%
\noindent
Let us start by focusing on the first term in the sum.
Because $(Y \indep Z|X,W)_{G_{\thickbar X,\thickubar Z}}$ rule 2 implies that
$$
P_x(y|w,z) = P_{x,z}(y|w)
$$
and by noting that $(Y \indep X|Z,W)_{G_{\thickbar Z,\thickbar X}}$ rule 3 allows us to write
$$
P_{x,z}(y|w) = P_z(y|w).
$$
By expanding the previous expression we get
\begin{equation} \label{eq:big2}
P_z(y|w) = \sum_x P_z(y|w,x)P_z(x|w).
\end{equation}
Rule 2 and the fact that $(Y \indep Z|X,W)_{G_{\thickubar Z}}$ together imply
\begin{equation} \label{eq:part11}
P_z(y|w,x) = P(y|w,x,z).
\end{equation}
The condition $(X \indep Z|W)_{G_{\thickbar Z}}$ and rule 3 allow us to write
\begin{equation} \label{eq:part12}
P_z(x|w) = P(x|w).
\end{equation}
Inserting (\ref{eq:part11}) and (\ref{eq:part12}) into (\ref{eq:big2}) yields
\begin{equation} \label{eq:bigsolve1}
P_z(y|w) = \sum_x P(y|w,x,z)P(x|w).
\end{equation}
Focusing now on the second term of (\ref{eq:big1}) we see that because $(Z \indep X|W)_{G_{\thickubar X}}$ rule 2 implies that
\begin{equation} \label{eq:part2}
P_x(z|w) = P(z|x,w).
\end{equation}
Similarly, the third term simplifies by using rule 3 and the condition $(W \indep X)_{G_{\thickbar X}}$ rule 3.
\begin{equation} \label{eq:part3}
P_x(w) = P(w).
\end{equation}
Finally, we combine the results above by inserting (\ref{eq:bigsolve1}), (\ref{eq:part2}) and (\ref{eq:part3}) into (\ref{eq:big1}) which yields the expression for the causal effect.
$$ P_x(y) = \sum_{w,z} \left( \sum_x P(y|w,x,z)P(x|w) \right) P(z|x,w)P(w) $$
In Section~\ref{Sect:practice} we will see how the causal effect can be identified by applying the algorithm of \citep{shpitser06}. The previous result highly resembles the front-door criterion, which states that
$$ P_{\+ x}(\+ y) = \sum_{\+ s} \left( \sum_{\+ x} P(\+ y|\+ x, \+s)P(\+ x) \right) P(\+ s|\+ x),$$
whenever the set $\+ S$ blocks all directed paths from $\+ X$ to $\+ Y$, there are no unblocked back-door paths from $\+ X$ to $\+ S$ and $\+ X$ blocks all back-door paths from $\+ S$ to $\+ Y$. However, neither $W$, $Z$, or $\{W,Z\}$ satisfy the role of the set $\+ S$. The criterion would certainly hold if we removed $W$ from the graph.

\section{Identifiability algorithm} \label{Sect:algo}
Even if a causal effect is identifiable, the rules of do-calculus themselves do not guarantee that they could be used to form an expression for the interventional distribution, and that it would contain only observed quantities. It is also not self-evident in which order the rules of do-calculus should be applied to reach the desired expression from the joint distribution of the observed variables $P(\+ V)$.

To overcome these limitations an identifiability algorithm has been developed by \citet{shpitser06}. This algorithm can be used to determine the identifiability of any causal effect, in addition of generating the expression for the interventional distribution in the case of an identifiable effect.

\subsection{Definitions} \label{Sect:definitions}

Some graph theoretic definitions are necessary in order to present the algorithm. The notation mostly follows that of \citep{shpitser06} with some slight alterations for the benefit of the reader.

\begin{definition}[Induced Subgraph] Let $H = \langle \+ W,\+ F \rangle$ and $G = \langle \+ V, \+ E \rangle $ be graphs such that $\+ W \subset \+ V$. If every pair of nodes $X,Y \in \+ W$ is connected by an edge in graph $H$ precisely when they are connected by an edge of the same direction in graph $G$, then $H$ is an \emph{induced subgraph} induced by the set $\+ W$ and $H = G[\+ W]$. 
\end{definition}

\begin{figure}[t!]
  \centering
  \vspace*{-0.5cm}
  \begin{subfigure}[b]{0.34\textwidth}
    \includegraphics[width=1\textwidth]{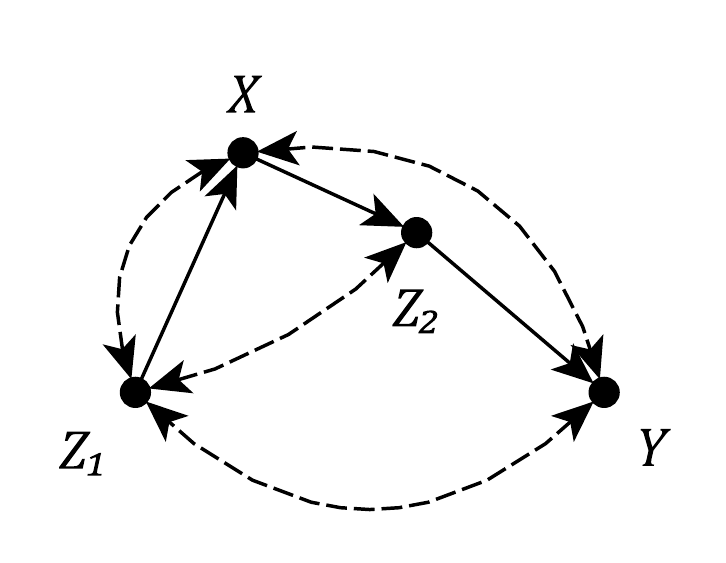}
    \caption{Graph $G$.}
    \label{kuva:grfGsubexpa}
  \end{subfigure}
  \begin{subfigure}[b]{0.50\textwidth}
    \hspace{1.5cm} \includegraphics[width=0.5\textwidth]{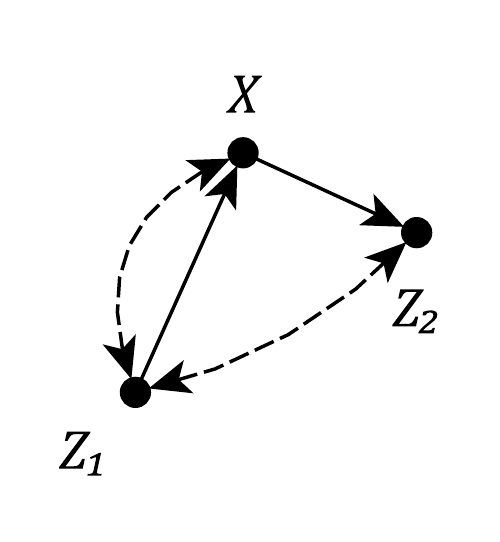}
    \caption{A subgraph of $G$ induced by the set $\{X,Z_1,Z_2\}$.}
    \label{kuva:grfGsubexpb}
  \end{subfigure}
  \caption{An example illustrating the definition of an induced subgraph.}
  \label{kuva:grfGsubexp}
\end{figure}

Defining new graphs using only a set of nodes can easily be achieved using induced subgraphs. For example, the graph in Figure \ref{kuva:grfGsubexp}(\subref{kuva:grfGsubexpb}) is an induced subgraph induced by the nodes $X, Z_1$ and $Z_2$ from  $G$ in \ref{kuva:grfGsubexp}(\subref{kuva:grfGsubexpa}).

\noindent
Perhaps the most important definition is C-component (confounded component). 

\begin{definition}[C-component, \citep{shpitser06} 3] Let $G = \langle \+ V,\+ E \rangle$ be a graph. If there exists a set $\+ B$ such that $\+ B \subset \+ E$ and $\+ B$ contains only bidirected edges, and the graph $\langle \+ V,\+ B \rangle$ is connected, then $G$ is a \emph{C-component}.
\end{definition}

Both graphs in Figure \ref{kuva:grfGsubexp} are examples of C-components. Even if a graph is not a C-component, at least one of its subgraphs is guaranteed to be a C-component because every subgraph induced by a single node is always a C-component. It is often of greater interest to determine how a given graph can be partitioned in C-components that contain as many nodes as possible.

\begin{definition}[Maximal C-component] Let $G$ be a graph and $C = \langle \+ V, \+ E \rangle$ a C-component such that $C \subset G$. C-component $C$ is \emph{maximal} (with respect to graph $G$) if $H \subset C$ for every bidirected path $H$ of graph $G$ which contains at least one node of the set $\+ V$.
\end{definition}

\citet{tian02phd} proved, that the joint probability distribution $P(\+ V)$ of the observed variables of graph $G$ can always be factorized in such a way, that each term of the resulting product corresponds to a maximal C-component. This property is in a fundamental role in the algorithm, since it can be used to recursively divide the expression of the interventional distribution into simpler expressions.

If a given graph $G$ is not a C-component, it can still be divided into a unique set $C(G)$ of subgraphs, each a maximal C-component of $G$. This follows from the fact, that there exists a bidirected path between two nodes in $G$ if and only if they belong in the same maximal C-component, which in turn follows from the definition of a maximal C-component. This means, that the bidirected paths of graph $G$ completely define its maximal C-components.

C-trees are a special case of C-components. They are closely related to direct effects, which are causal effects of the form $P_{Pa(Y)}(Y)$.

\begin{definition}[C-tree, \citep{shpitser06} 4] Let $G$ be a C-component such that every observed node has at most one child. If there is a node $Y$ such that $G[An(Y)_G] = G$, then $G$ is a $Y$-\emph{rooted C-tree}.
\end{definition}
\noindent
Using only C-trees and C-components it is already possible to characterize identifiability of effects on a single variable. C-forest is the multivariate generalization of a C-tree in such a way that the \emph{root set}, which is the set of nodes $\{X \in G\mid De(X)_G\setminus\{X\} = \emptyset\}$, contains one or more nodes.

\begin{definition}[C-forest, \citep{shpitser06} 5] Let $G$ be a graph and $\+ Y$ its root set. If $G$ is a C-component, and every observed node has at most one child, then $G$ is $\+ Y$-\emph{rooted C-forest}.
\end{definition}

Both C-components in Figure \ref{kuva:grfGsubexp} are also C-forests, because every observed node has at most one child in both graphs. In addition, their root sets consist only of a single node. There exists a connection between C-forests and general causal effects of the form $P_{\+ x}(\+ Y)$. A graph structure formed by a pair of C-trees is used to determine such effects.

\begin{figure}[t!]
	\centering
	\includegraphics[width=0.52\textwidth, trim=0 13 0 13, clip]{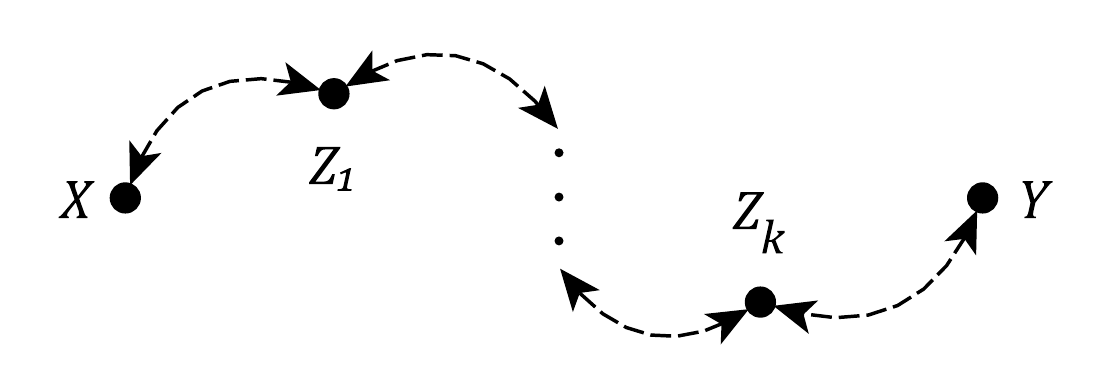}
  \caption{Path $H$.}
  \label{kuva:grfGzbpath}
\end{figure}
\citet{shpitser06} proved, that if a graph $G$ contains a hedge for $P_{\+ x}(\+ y)$, then the effect is not identifiable.

\begin{definition}[Hedge, \citep{shpitser06} 6] Let $G = \langle \+ V, \+ E \rangle $ be a graph, and $\+ X, \+ Y \subset \+ V$ disjoint subsets. If there are two $\+ R$-rooted C-forests $F = \langle \+ V_F, \+ E_F \rangle$ and $F^\prime = \langle \+ V_{F^\prime}, \+ E_{F^\prime} \rangle$ such that ${\+ V_F \cap \+ X \neq \emptyset}$, $\+ V_{F^\prime} \cap \+ X = \emptyset, F^\prime \subset F,$ and $\+ R \subset An(\+ Y)_{G_{\thickbar {\+ X}}}$, then $F$ and $F^\prime$ form \emph{hedge} for $P_{\+ x}(\+ y)$ in $G$.
\end{definition}
Hedges are a remarkable structure, since they generalize certain results regarding identifiability. One example of such a result is the condition for identification of a causal effect of the form $P_x(\+ y)$ in \citep{tian02}. The result states that $P_x(\+ y)$ is identifiable if and only if there are no bidirected paths between $X$ and any of its children in $G[An(\+ Y)_G]$. Consider the graph $H = \langle \+ V, \+ E \rangle$ in Figure \ref{kuva:grfGzbpath} containing the nodes $X$ and $Y$ and a bidirected path connecting them formed by the intermediary nodes $\{Z_1,\ldots,Z_k\}$. One can observe, that the C-forests $H$ and $H[\+ V\setminus\{X\}]$ form a hedge for $P_x(Y, Z_1, \ldots, Z_k)$.

\subsection{Algorithm}\label{Sect:algos}

Using the previously presented definitions it is now possible to define Algorithm~\ref{alg:identify}, which completely characterizes the identifiability problem of general causal effects. \citet{shpitser06} showed, that the expression returned by Algorithm~\ref{alg:identify} for $P_{\+ x}(\+ y)$ is always correct if the effect in question is identifiable. They also showed, that if the algorithm is interrupted on line five, then the original graph $G$ contains a hedge, preventing the identifiability of the effect. The existence of a hedge is therefore equivalent with unidentifiability. This result also shows the completeness of do-calculus, because the algorithm only applies standard rules of probability manipulations and the three rules of do-calculus. All variables are assumed to be discrete, but the algorithm can also be applied in a continuous case, when the respective sums are replaced with integrals.

The algorithm is required to be able to iteratively process the nodes of the graph, which means that the nodes have to be ordered in some meaningful fashion. This ordering must be able to take the directions of the edges into account, and at least one such ordering must always exist for any given graph. Topological ordering has all of these prerequisite properties.

\begin{definition}[Topological Ordering]
\emph{Topological ordering} $\pi$ of a DAG $G = \langle \+ V, \+ E \rangle$ is an ordering of its nodes, where either $X > Y$ or $Y > X$ for all pairs of nodes $X,Y \in \+ V,\, X \neq Y$ in $G$. In addition, no node can be greater than its descendants in $\pi$. In other words, if $X$ is an ancestor of $Y$ in $G$, then $X < Y$. 
\end{definition}

There exists at least one topological ordering for any DAG, but in some cases there can be multiple orderings. One way to always construct an ordering for a given graph is to begin by determining all nodes without parents, and ordering them arbitrarily. Next, all nodes without parents excluding the nodes found in previous step are determined and again ordered arbitrarily. It is also assigned, that the largest node in the previous step is smaller than the smallest node in the current step. This process is iterated, until all nodes have been ordered.

\begin{algorithm}[t!]
	\begin{algorithmic}[1]
		\INPUT{Value assignments $\+ x$ and $\+ y$, joint distribution $P(\+ v)$ and a DAG $G = \langle \+ V, \+ E \rangle$. $G$ is an $I$-map of $P$.}
		\OUTPUT{Expression for $P_{\+ x}(\+ y)$ in terms of $P(\+ v)$ or \textbf{FAIL}$(F,F^\prime)$.}
		\Statex
		\Statex \textbf{function}{ \textbf{ID}$(\+ y, \+ x, P, G)$}
			\If{$\+ x = \emptyset$,}
				\Statex \textbf{\quad return}{ $\sum_{v \in \+ v \setminus \+ y}P(\+ v)$.}
			\EndIf
			\If{$\+ V \neq An(\+ Y)_G,$}
				\Statex \textbf{\quad return}{ \textbf{ID}$(\+ y, \+ x \cap An(\+ Y)_G, P(An(\+ Y)_G), G[An(\+ Y)_G)]$.}
			\EndIf	
			\State \textbf{Let}{ $\+ W = (\+ V \setminus \+ X) \setminus An(\+ Y)_{G_{\thickbar {\+ X}}}$.}
			\Statex \textbf{if}{ $\+ W \neq \emptyset $,} \textbf{then}
				\Statex \quad \textbf{return}{ \textbf{ID}$(\+ y, \+ x \cup \+ w, P, G)$.}
			\If{$C(G[\+ V \setminus \+ X]) = \{G[\+ S_1], \ldots,G[\+ S_k]\}$,}
				\Statex \textbf{\quad return}{ $\sum_{v \in \+ v \setminus (\+ y \cup \+ x)} \prod_{i=1}^k$ \textbf{ID}($\+ s_i, \+ v \setminus \+ s_i, P, G)$.}
			\EndIf
			\Statex \textbf{if}{ $C(G[\+V \setminus \+ X]) = \{G[\+ S]\}$,} \textbf{then}
				\State \quad \textbf{if}{ $C(G) = \{G\}$,} \textbf{then}
				\Statex \quad \quad \textbf{throw FAIL}{$(G, G[\+ S])$.}
				\State \quad \textbf{if}{ $G[\+ S] \in C(G)$,} \textbf{then} 
				\Statex \quad \quad \textbf{return} {$\sum_{v \in \+ s \setminus \+ y} \prod_{V_i \in \+ S}{P(v_i \vert v_\pi^{(i-1)})}$.}
				\State \quad \textbf{if}{ $(\exists \+ S^\prime)\+ S \subset \+ S^\prime \text{ such that } G[\+ S^\prime] \in C(G)$,} \textbf{then}
				\Statex \quad \quad \textbf{return} {\textbf{ID}$(\+ y, \+ x \cap \+ s^\prime, \prod_{V_i \in \+ S^\prime}{P(V_i \vert V_\pi^{(i-1)} \cap \+ S^\prime,v_\pi^{(i-1)} \setminus \+ s^\prime), G[\+ S^\prime}])$.}
	\end{algorithmic}
	\caption{The causal effect of intervention $do(\+ X = \+ x)$ on $\+ Y$.}
	\label{alg:identify}
\end{algorithm}
Algorithm~\ref{alg:identify} is simple in a sense that at each recursion stage the computation proceeds to exactly one line only. This is easy to see from the fact that after a condition regarding any of the line has been checked, either a \textbf{return} or a \textbf{FAIL} command will be executed. If $\+ x = \emptyset$ on line one, then the marginal distribution $P(\+ y)$ is computed instead of a causal effect. This can be achieved by marginalizing over the joint distribution $P(\+ V)$. On line two, all non-ancestors of $\+ Y$ in $G$ are eliminated. This is possible due to the fact that the input of the algorithm assumes that $G$ is an $I$-map of $G$ and thus all necessary conditional independences hold. On line three, interventions are added to the original causal effect, which is feasible due to the third rule of do-calculus, because $(\+ Y \indep \+ W|\+X)_{G_{\thickbar {\+ X},\thickbar{\+ W}}}$.

It is possible to index the nodes of $G$ and the nodes of any subgraph of $G$ using the topological ordering. This property is utilized on lines four, six and seven. The notation $V_\pi^{(i-1)}$ refers to all nodes in $G$ that are smaller than $V_i$ in $\pi$. Any topological ordering of $G$ is also a topological ordering for any subgraph of $G$. This means, that it is unnecessary to determine a new ordering for each subgraph of $G$. Instead, one can fix the ordering before applying the algorithm.

The maximal C-components of $G[\+ V \setminus \+ X]$ are determined on line four and their factorization property is utilized. If more than one C-components were found, it is now necessary to calculate a new causal effect for every C-component. The algorithm proceeds to either line five, six or seven in the case if only one C-component was found.

If Algorithm~\ref{alg:identify} throws \textbf{FAIL}, then the original graph $G$ contains a hedge formed by graph $G$ and $G[\+ S]$ of the current recursion stage, due to which the original effect is not identifiable and computation terminates. If the algorithm continues, then it is necessary to determine whether $G[\+ S]$ is a maximal C-component of $G$. If this is the case, then the condition of line six has been satisfied. In the other case, the computation of the intervention can be limited to the intersection of sets $\+ X$ and $\+ S^\prime$ on line seven.

Identifiability of conditional interventional distributions is characterized by Algorithm~\ref{alg:identifycond}. This algorithm is a generalization of Algorithm~\ref{alg:identify} and in fact it utilizes the function \textbf{ID} in the computation. It was constructed by \citet{shpitser2006identificationconditional} for identifying conditional causal effects i.e., causal effects of the form $P_{\+ x}(\+ y|\+ z)$. They showed, that this algorithm is also sound and complete for identifying all such effects.

\begin{algorithm}[t!]
	\begin{algorithmic}[1]
		\INPUT{Value assignments $\+ x$, $\+ y$ and $\+ z$, joint distribution $P(\+ v)$ and a DAG $G = \langle \+ V, \+ E \rangle$. $G$ is an $I$-map of $P$.}
		\OUTPUT{Expression for $P_{\+ x}(\+ y| \+ z)$ in terms of $P(\+ v)$ or \textbf{FAIL}$(F,F^\prime)$.}
		\Statex
		\Statex \textbf{function}{ \textbf{IDC}$(\+ y, \+ x,\+ z, P, G)$}
			\If{$\exists Z \in \+ Z \text{ such that } (\+ Y \indep Z|\+ X, \+ Z \setminus \{Z\})_{G_{\thickbar {\+ X},\thickubar {\+ Z}}}$}
				\Statex \textbf{\quad return} { \textbf{IDC}$(\+ y, \+ x \cup \{z\}, \+ z \setminus \{z\}, P, G)$.}
			\EndIf
			\State \textbf{else let} { $P^\prime =\;$\textbf{ID}$(\+ y \cup \+ z, \+ x, P, G)$.}
			\Statex \textbf{return} {$P^\prime / \sum_{y \in \+ y}P^\prime$}
	\end{algorithmic}
	\caption{The causal effect of intervention $do(\+ X = \+ x)$ on $\+ Y$ given $\+ Z$.}
	\label{alg:identifycond}
\end{algorithm}

The primary focus of this paper however, is the implementation of Algorithm~\ref{alg:identify}. The implementation of Algorithm~\ref{alg:identifycond} follows seamlessly from this implementation, because at the bottom of any recursive stack of \textbf{IDC} the function \textbf{ID} is ultimately called, which determines if the original conditional effect is identifiable. The only additional task is to determine whether a suitable node for the d-separation condition exists on line 1.

\subsection{Application in practice} \label{Sect:practice}

\begin{figure}[t!]
  \centering
  \begin{subfigure}[b]{0.28\textwidth}
    \includegraphics[width=1\textwidth]{g1.pdf}
    \caption{Graph $G$}
    \label{kuva:grfGa}
  \end{subfigure}
  \qquad
  \begin{subfigure}[b]{0.29\textwidth}
    \includegraphics[width=1\textwidth]{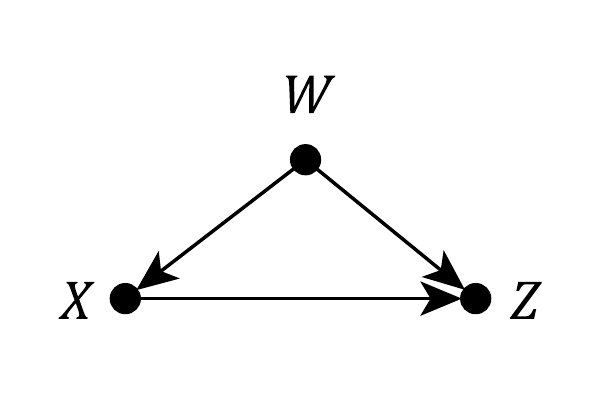}
    \caption{Subgraph $G[An(Z)_G]$.}
    \label{kuva:grfGb}
  \end{subfigure}
  \qquad
  \begin{subfigure}[b]{0.28\textwidth}
    \includegraphics[width=1\textwidth]{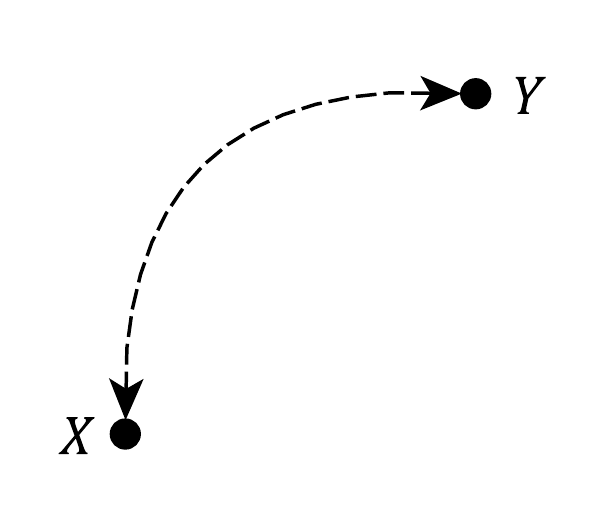}
    \caption{Subgraph  $G[\+ S^\prime]$.}
    \label{kuva:grfGc}
  \end{subfigure}
  \caption{Graph $G$ and its subgraphs.}
  \label{kuva:grfG}
\end{figure}

We return to the example presented in Section~\ref{Sect:docalc}. The graph of the example along with some subgraphs are shown here in Figure~\ref{kuva:grfG}. Let $G = \langle \+ V, \+ E \rangle$ be a graph such as in Figure \ref{kuva:grfG}(\subref{kuva:grfGa}) and a causal effect of interest $P_x(y)$, which is to be identified from the joint distribution $P(X, Y, Z, W)$. Only a single topological ordering exists for the nodes of $G$, and it is $W<X<Z<Y$. Clearly $\+ x \neq \emptyset, \+ V = An(Y)_G$ and $\+ W = \emptyset$, so the first three lines are ignored and line four is triggered, since
$$ C(G[\+ V \setminus \{X\}]) = \{G[W], G[Z], G[Y]\}.$$ 
Because $\+ v \setminus (\{y\} \cup \{x\}) = \{w, z\}$, it is now necessary to identify three new causal effects in the following expression:
$$ \sum_{w, z}P_{x, z, y}(w)P_{w, x, y}(z)P_{w, x, z}(y).$$ 
Consider the first term of the product. Because $\+ V \neq An(W)_G$, line two is triggered, and non-ancestors of $W$ are ignored. 
\noindent
This results in the first term simplifying to $P(w)$ because $An(W)_G = \{W\}$. Line two is also triggered when computing the second term, and 
$$ P_{w,x,y}(z) = P_{w,x}(z)$$ 
in a subgraph induced by ancestors of $Z$ as in Figure \ref{kuva:grfG}(\subref{kuva:grfGb}). Observing that 
$$ C(G[An(Z)_G \setminus \{W,X\}]) = \{G[Z]\}$$ 
and
$$ G[Z] \in C(G[An(Z)_G]) = \{G[X],G[W],G[Z]\},$$
the algorithm proceeds to line 6 and the second term simplifies again
$$ P_{w,x}(z) = P(z \vert w,x).$$
The last term $P_{w,x,z}(y)$ triggers line four, because
$$ C(G[\+ V \setminus \{W,X,Z\}]) = \{G[Y]\}.$$ 
$G[Y]$ is not a maximal C-component of $G$, but $Y$ is a node of one of the maximal C-components of $G$:
$\{Y\} \subset \{X,Y\} = \+ S^\prime$. It holds for the set $\+ S^\prime$, that 
$$ G[\+ S^\prime] \in C(G) = \{G[\{X,Y\}],G[W],G[Z]\}.$$ 
So it is mandatory to compute $P_x(y)$ from $P(X|w)P(Y \vert X,w,z)$ in the graph corresponding to Figure \ref{kuva:grfG}(\subref{kuva:grfGc}). It should be noted, that this causal effect differs from the original effect $P_x(y)$, because the joint distribution $P(\+ V)$ of observed variables of $G$ is not the same as the distribution $P(X|w)P(Y \vert X,w,z)$ of the subgraph of the current recursion stage. 

Line two is triggered next, and since $Y$ has no observed ancestors in the graph corresponding to \ref{kuva:grfG}(\subref{kuva:grfGc}), it follows that
$$P_x(y) = \sum_x P(x|w)P(y \vert x,w,z).$$ 
An expression for the original causal effect is obtained by combining the previous results
$$ P_x(y) = \sum_{w,z}P(z\vert w,x)P(w)\sum_{x}P(y\vert w,x,z)P(x\vert w). $$
The result agrees with the result derived in Section~\ref{Sect:docalc}.

\begin{figure}[t!]
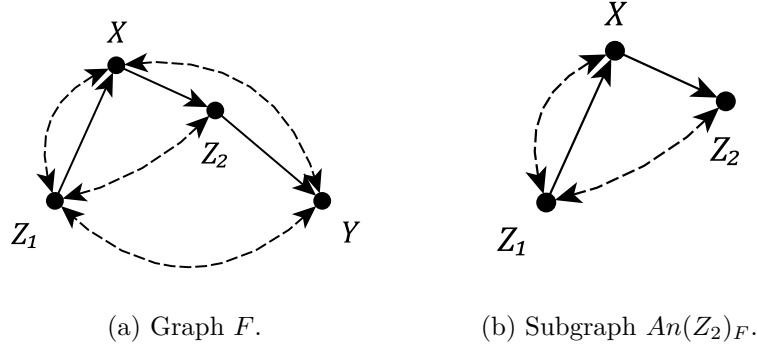

  \vspace*{-0.5cm}
	\centering
	\begin{subfigure}[b]{0.35\textwidth}
		\includegraphics[width=1\textwidth]{g2.pdf}
    \caption{Graph $F$.}
		\label{kuva:grfFa}
	\end{subfigure}
	\qquad
	\begin{subfigure}[b]{0.27\textwidth}
		\includegraphics[width=1\textwidth]{g2sub1.pdf}
    \caption{Subgraph $An(Z_2)_F$.}
		\label{kuva:grfFb}
	\end{subfigure}
	\caption{Graph $F$ and its subgraph $F[An(Z_2)_F]$.}
  \label{kuva:grfF}
\end{figure}

Algorithm~\ref{alg:identify} can also be used to detect unidentifiability. Let $F = \langle \+ V, \+ E \rangle$ be a graph of Figure \ref{kuva:grfF}(\subref{kuva:grfFa}) and a causal effect of interest $P_x(y)$, which is to be identified from $P(X,Y,Z_1,Z_2)$. Let the topological ordering of the nodes of $F$ be $Z_1 < X < Z_2 < Y$.

\noindent
The computation starts from line three 
$$\+ W = (\+ V \setminus \+ X) \setminus An(\+ Y)_{F_{\thickbar {\+ X}}} = (\{X,Y,Z_1,Z_2\} \setminus \{X\}) \setminus \{X,Z_2,Y\} = \{Z_1\} \neq \emptyset.$$ 
$Z_1$ is added to the original intervention, so $P_{{z_1},x}(y)$ has to be identified. Line four is triggered next, because 
$$C(F[\+ V \setminus \{Z_1,X\}]) = \{F[Z_2],F[Y]\}.$$ 
Since $\+ v\setminus (\{y\}\cup\{z_1,x\}) = \{z_2\}$, two new causal effects have to be identified following expression:
$$\sum_{z_2}P_{z_1,x,y}(z_2)P_{z_1,x,z_2}(y).$$
Consider the first term of the product. Clearly $\+ V \neq An(Z_2)_F$, so the algorithm proceeds to line two, which means that 
$$P_{z_1,x,y}(z_2) = P_{z_1,x}(z_2)$$ in a subgraph formed by ancestors of $Z_2$ as in Figure \ref{kuva:grfF}(\subref{kuva:grfFb}). However, $P_{z_1,x}(z_2)$ is not identifiable, because
$$C(F[An(Z_2)_F\setminus \{Z_1,X\}]) = \{F[Z_2]\} \;\text{ and }\; C(F[An(Z_2)_F]) = \{F[An(Z_2)_F]\},$$
which trigger line five. 
In conclusion, $F$ contains a hedge for $P_{z_1,x}(z_2)$ formed by C-forests $F[Z_2]$ and $F[\{Z_1,Z_2,X\}]$. Thus the original effect $P_x(y)$ is not identifiable.

\section[Implementation using R]{Implementation using \proglang{R}} \label{Sect:implementinR}

The programming language \proglang{R} \citep{rsoft} was chosen for the implementation of Algorithm~\ref{alg:identify}. The \proglang{R} packages \pkg{XML} \citep{xml}, \pkg{igraph} \citep{igraph} and \pkg{ggm} \citep{ggm} are utilized repeatedly throughout the implementation. 

\subsection{Graph files} \label{Sect:files}
A graph $G$ induced by the causal model is a crucial argument of Algorithm~\ref{alg:identify}. Many file formats for visualizing graphs are available, each with their own strengths and weaknesses. Some of these formats are very simple, and do not differentiate directed and undirected graphs. Some formats offer excessive features for describing causal models, or they might require handling complex syntax, which can be time consuming.

GraphML \citep{graphml} is a user-friendly file format for graphs. Its features include support for directed graphs and visualizations. GraphML is based on the extensible markup language XML \citep{Maler:04:EML}, which makes processing of graphs files almost effortless. One can also include the names of the nodes within the GraphML file itself, so the user is not limited to having to input the node names themselves inside the \proglang{R} environment. Graphical editors for creating GraphML files are freely available for the user. A special function called \code{parse.graphml} has been developed for processing GraphML files. However, the implementation of Algorithm~\ref{alg:identify} is not limited to GraphML files alone. Any file format supported by the \pkg{igraph} package can be used, as long as the graph follows one of the following notations for bidirected edges. 

Bidirected edges can be separated from unidirected edges by using graphical parameters. For this purpose, three distinct notations have been selected to describe bidirected edges, which correspond to unobserved nodes.

\begin{figure}[t!]
  \vspace*{-0.35cm}
	\centering
	\begin{subfigure}[b]{0.28\textwidth}
		\includegraphics[width=1\textwidth]{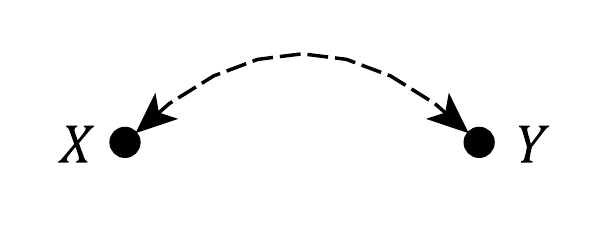}
    \caption{Notation 1.}
		\label{kuva:bidira}
	\end{subfigure}
	\qquad
	\begin{subfigure}[b]{0.28\textwidth}
		\includegraphics[width=1\textwidth]{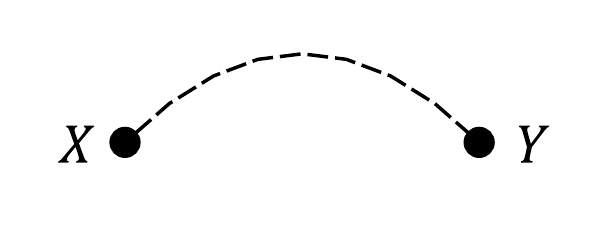}
        	\caption{Notation 2.}
		\label{kuva:bidirb}
	\end{subfigure}
	\qquad
	\begin{subfigure}[b]{0.28\textwidth}
		\includegraphics[width=1\textwidth]{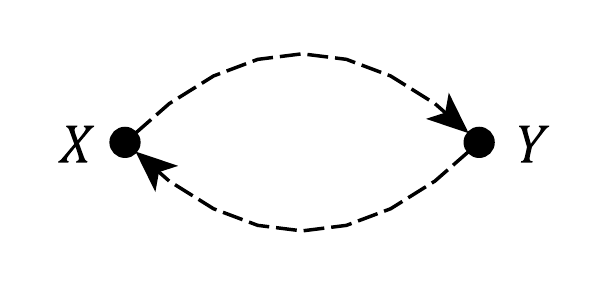}
    \caption{Notation 3.}
		\label{kuva:bidirc}
	\end{subfigure}
  \caption{Notations for bidirected edges.}
  \label{kuva:bidir}
\end{figure}
\noindent
The available notations for bidirected edges are shown in Figures \ref{kuva:bidir}(\subref{kuva:bidira}), \ref{kuva:bidir}(\subref{kuva:bidirb}) and \ref{kuva:bidir}(\subref{kuva:bidirc}). It should be noted, that notations 1 and 2 are almost identical. Because of their similarity, both notations 1 and 2 are referred to as \code{standard} notation. Notation 3, as shown in Figure~\ref{kuva:bidir}(\subref{kuva:bidirc}) differs from the previous two. It is apparent, that this notation cannot be used as such, because it induces loops in the graph which is not allowed in the context of DAGs. However, GraphML format enables the assignment of parameters for the edges, which in turn allows one to separate these edges from their unidirected counterparts. When using notation 3, one must define a parameter called \code{description} for the two unidirected edges corresponding to the bidirected edge, and assign its value to \code{"U"} (Unobserved). Notation 3 is used in the implementation itself, which is why it is referred to as the \code{internal} notation.

The process of importing GraphML files created by a graphical editor is handled by using the \proglang{R} package \pkg{XML}. This package contains the function \code{xmlParse}, which is utilized to import graph files into \proglang{R} objects. It should be noted, that these objects only reflect their internal \proglang{C} objects and are thus different from ordinary \proglang{R} objects. This means that the memory reserved by the XML objects has to be freed after the files have been imported. Normally \proglang{R} does this automatically.

Algorithm~\ref{alg:identify} requires only a small portion of the XML content, and the unnecessary content is removed in the process of searching for the important items. Items of importance are those that contain data about the node names, node count, edge count and the values of the \code{description} parameters of the edges. If notation 1 or 2 of Figures \ref{kuva:bidir}(\subref{kuva:bidira}) and \ref{kuva:bidir}(\subref{kuva:bidirb}) was used for the bidirected arcs, it is converted to match the \code{internal} format of Figure \ref{kuva:bidir}(\subref{kuva:bidirc}). The XML search is implemented using the function \code{getNodeSet} of the \pkg{XML} package. This function uses \proglang{XPath}, which is a processing language for XML content search \citep{simpson02}.

When the crucial information has been extracted, an \pkg{igraph} graph is formed from the remaining content. \pkg{igraph} is a tool for visualizing and processing graphs, and it can handle graphs which may contain millions of nodes due to its implementation in \proglang{C}. This package also offers many useful functions related to Algorithm~\ref{alg:identify}, such as determining the ancestors of a node, constructing a topological ordering and generating induced subgraphs from a set of edges or nodes. One of the main goals of \pkg{igraph} is the effortless implementation of graph algorithms.

\subsection{Distribution objects} \label{Sect:distobjs}
An important question regarding the Algorithm~\ref{alg:identify} of Section~\ref{Sect:algos}, is how the probability distribution which changes at each recursive stage should be implemented. An intuitive solution is to construct a distribution object, which maintains the terms currently present in the expression. Distribution objects are recursive by construction as is the algorithm itself. In practice this means that when any of the lines four, six or seven is triggered, sub objects are formed, which correspond to the product terms of the expression. These sub objects can further branch into sub objects of their own and so forth. \pkg{causaleffect} implements an \proglang{R} class called \code{probability} to represent the distribution objects.

Multiple attributes have to be set for the distribution objects in order to present the probability distribution precisely. The string vectors \code{var} and \code{cond} are one of the most common attributes, because they enable the definition of a simple conditional distribution. A distribution is formed by the variables described in \code{var} conditioned on those of \code{cond}. For example, let \code{p} be a distribution object, and let the values of its attributes be \code{var = "Y"} and \code{cond = "X"}. Therefore object \code{p} represents the conditional distribution $P(Y | X).$

When the distribution is a product, the individual terms are defined in a list of distribution objects called \code{children} and a logical variable \code{recursive} is set to \code{TRUE} to differentiate this object from those containing only a single term. For example, for a distribution which represents the distribution $P^* = P(Z \vert X)P(X|Y)P(Y)$ one has to set \code{children = list(a,b,c)}, where the objects \code{a}, \code{b} and \code{c} represent the distributions $P(Z \vert X), P(X|Y)$ and $P(Y)$ respectively.

For marginal distributions a string vector \code{sumset} has been defined. The contents of this vector correspond to the variables which the distribution is to be summed over in the discrete case, or integrated over in the continuous case. In simple situations this parameter is not needed, but often with more complex graphs one encounters instances, where the computation of conditionals is no longer straightorward. Suppose one had to compute the marginal distribution $P^*(X)$ of $X$ from the joint distribution $P^*(X,Y,Z)$ of the previous example. To achieve this, one has to set \code{sumset = c("Y","Z")} for the matching distribution object, because $P^*(X) = \sum_{Y,Z}P(Z \vert X)P(X|Y)P(Y)$.

The level of complexity increases further when computing conditionals from distributions which consist of multiple product terms. The previously presented attributes are often insufficient to form an expression for the corresponding distribution object. Consider once more the joint distribution $P^*$. Computing the marginal conditional distribution $P^*(X | Y)$ results in
\begin{align*}
 P^*(X \vert Y) = \frac{P^*(X,Y)}{P^*(Y)} = \frac{\sum_{Z}P(Z \vert X)P(X|Y)P(Y)}{\sum_{X,Z}P(Z \vert X)P(X|Y)P(Y)} &= \\ \frac{P(X|Y)\sum_{Z}P(Z \vert X)}{\sum_{X}P(X|Y)\sum_{Z}P(Z \vert X)} &= P(X|Y).
\end{align*}
The implementation is able to handle similar situations, where the expression can easily be simplified using the following procedure. Any term which does not depend on the summation index, will be placed outside of the sum. Next, it is checked whether any expressions can be simplified by changing the order of summation. Corresponding terms are subtracted if possible.

These simplification rules are not sufficient to handle every situation. For example, the expression $\sum_{X}P(Y \vert X)P(X)$ cannot be simplified using the procedure above. One cannot remove any terms from within the sum and the summation order is clearly fixed. In situations, where the denominator is necessary in order to correctly form the expression, one needs to include additional attributes called \code{divisor} and \code{fraction}. These attributes are similar to the attributes \code{children} and \code{recursive} in a sense that \code{divisor} contains the distribution object that represents the denominator and \code{fraction} is set to \code{TRUE} when it is necessary to represent the expression as a fraction.

\subsection{Maximal C-components} \label{Sect:ccomps}
In Section~\ref{Sect:definitions} it was shown, that for every causal diagram $G$ there exists a unique set $C(G)$ of maximal C-components of $G$. To construct this set, one has to begin by determining all bidirected edges of $G$. Afterwards, a subgraph containing only bidirected edges is formed. This subgraph will contain one or more \emph{components}, which are connected subgraphs of $G$. Because these components are disjoint and every pair of nodes within a component is connected by a bidirected path, it follows that they must be the maximal C-components of $G$. The \emph{adjacency matrix} of $G$ is utilized to find the bidirected edges of $G$.

\begin{definition}[adjacency matrix]
An \emph{adjacency matrix} of a graph $G = \langle \+ V, \+ E \rangle$ is a ${n \times n}$ matrix $A = [a_{ij}]$, where $n$ is the number of nodes of $G$, $\+ V = \{V_1, V_2, \ldots, V_n\}$ and $a_{ij}$ is the number of edges from $V_i$ to $V_j$. 
\end{definition}

Because $G$ is directed, its adjacency matrix is not necessarily symmetric. When notation 3 of Figure \ref{kuva:bidir}(\subref{kuva:bidirc}) is used to describe the bidirected edges, it is easy to confirm that two nodes $V_i$ and $V_j$ are connected by at least one bidirected edge if and only if $a_{ij} \geq 1$ and $a_{ji} \geq 1$. Thus all bidirected edges can be determined by comparing $A$ to its transpose $A^\top$, and by choosing only those edges which correspond to indices with $a_{ij} \geq 1$ and $a_{ji} \geq 1$.

The subgraph of $G$ containing only bidirected edges is constructed by using the function \code{subgraph.edges} of the \pkg{igraph} package. This function retains all nodes of the input graph, but removes all the edges that were not given as input. The subgraph returned by this function is further divided into components by using the function \code{decompose.graph} which is also provided by \pkg{igraph}.

\subsection{Implementation} \label{Sect:implement}
All necessary preparations have been presented to implement Algorithm~\ref{alg:identify}. Any probability distribution can be represented with a corresponding distribution object, and the adjacency matrix provides a method to determine the maximal C-components of $G$. Other important methods are provided by the \pkg{igraph} package, such as constructing subgraphs and determining the ancestors of a given set of nodes. In this implementation, the input of Algorithm~\ref{alg:identify} consists of the sets $\+ x$ and $\+ y$ including the graph $G$, and returns a \code{probability} object, which is a list structure that describes the expression of the causal distribution $P_{\+ x}(\+ y)$ in terms of $P(\+ V)$. The returned object can be further parsed into a character representation.
 
The \proglang{R} function of Algorithm~\ref{alg:identify} is called \code{id}. This function takes five parameters as input: a string vector \code{y}, a string vector \code{x}, a distribution object \code{P}, an \pkg{igraph} graph \code{G} and a string vector \code{to}. The first four parameters correspond to their mathematical counterparts, namely the vectors $\+ x$, $\+ y$, $P$ and $G$. The last parameter \code{to} is a string vector representing some topological ordering of the nodes of $G$. All required set theoretic operations are included in \proglang{R} as the functions \code{intersect}, \code{setdiff} and \code{union}. 

The observed portion of \code{G} is saved as \code{G.obs}. This graph contains all the observed nodes of $G$ and the edges between them. In addition, the observed nodes are saved into vector \code{v}, and the ancestors of \code{y} are saved into vector \code{anc}. The implementation of each line of Algorithm~\ref{alg:identify} is presented next.

\vspace{0.1cm}
\begin{algorithmic}[1]
	\If{$\+ x = \emptyset$,}
		\Statex \textbf{\quad return}{ $\sum_{v \in \+ v \setminus \+ y}P(\+ v)$.}
	\EndIf
\end{algorithmic}
The truth value of the expression $\+ x = \emptyset$ is determined on line 1. This is done by computing the length of \code{x}. If the length is zero, then \code{id} combines the difference of the sets \code{v} and \code{y} with the \code{sumset} of \code{P} and returns \code{P}.

\vspace{0.1cm}
\begin{algorithmic}[1]
\setcounter{ALG@line}{1}
	\If{$\+ V \neq An(\+ Y)_G,$}
		\Statex \textbf{\quad return}{ \textbf{ID}$(\+ y, \+ x \cap An(\+ Y)_G, P(An(\+ Y)_G), G[An(\+ Y)_G)]$.}
	\EndIf	
\end{algorithmic}
The truth value of the condition on line 2 is determined by computing the length of the vector \code{setdiff(v, anc)}.
If the length is not zero, then \code{id} is called with the arguments \code{id(y, intersect(x, anc), P, anc.graph, to)}, where \code{anc.graph} is the induced subgraph $G[An(\+ Y)_G]$, which is constructed by using the \texttt{induced.subgraph} function of the \pkg{igraph} package. This function takes a set of nodes and a graph as input, and constructs a subgraph, which retains all of the nodes given as input, and all of the edges between them in the original graph.
\vspace{0.1cm}
\begin{algorithmic}[1]
\setcounter{ALG@line}{2}
	\State \textbf{let}{ $\+ W = (\+ V \setminus \+ X) \setminus An(\+ Y)_{G_{\thickbar {\+ X}}}$.}
	\Statex \textbf{if}{ $\+ W \neq \emptyset $,} \textbf{then}
		\Statex \quad \textbf{return}{ \textbf{ID}$(\+ y, \+ x \cup \+ w, P, G)$.}
\end{algorithmic}
To construct a vector \code{w} which represents the node set $\+ W$, one must first construct the subgraph $G_{\thickbar {\+ X}}$. To accomplish this, all incoming edges of $\+ X$ have to be determined. A useful operator is provided by the \pkg{igraph} package to accomplish this. The operator \code{\%->\%} can be used to find incoming or outgoing edges of a node. In this case, one finds the incoming nodes of \code{x} with the command \code{E(G) [1:length(E(G)) \%->\% x]}, where \code{E} is a function that returns all edges of \code{G}. When the subgraph has been constructed, \code{w} can also be constructed. If the length of \code{w} is not zero, then \code{id} is called with the arguments \code{id(y, union(x, w), P, G, to)}.

\vspace{0.1cm}
\begin{algorithmic}[1]
\setcounter{ALG@line}{3}
	\If{$C(G[\+ V \setminus \+ X]) = \{G[\+ S_1],\ldots,G[\+ S_k]\}$,}
		\Statex \textbf{\quad return}{ $\sum_{v \in \+ v \setminus (\+ y \cup \+ x)} \prod_{i=1}^k$ \textbf{ID}($\+ s_i, \+ v \setminus \+ s_i, P, G)$.}
	\EndIf
\end{algorithmic}
The set $C(G[\+ V \setminus \+ X])$ can be found with the function \texttt{c.components}. This function determines the node set of every maximal C-component of the input graph, and returns them as a list \code{s}. If the length of this list is larger than one, then \code{id} returns a new distribution object with \code{sumset = setdiff(v, union(y, x)), recursive = TRUE, children = productlist}, where every object in \code{productlist} is determined by a new recursive call for every C-component $G[\+ S_i],\, i = 1,\ldots,k$ that was found. These components are constructed by calling \code{id} with the arguments \code{id(s[[i]], setdiff(v, s[[i]]), P, G, to)}, $i = 1,\ldots,k$.

If the algorithm did not proceed to any of the previous lines, then the additional condition $C(G[\+ V \setminus \+ X]) = \{G[\+ S]\}$ must be true. The node set of the single C-component $G[\+ S]$ is now saved in the vector \code{s}, which was previously a list. This means that \code{s} is replaced by \code{s[[1]]}.

\vspace{0.1cm}
\begin{algorithmic}[1]
\setcounter{ALG@line}{4}
	\State \quad \textbf{if}{ $C(G) = \{G\}$,} \textbf{then}
		\Statex \quad \quad \textbf{throw FAIL}{$(G, G[\+ S])$.}
\end{algorithmic}
The function \code{c.components} is utilized again in order to find the maximal C-components of $G$.
If in addition to having only a single C-component this C-component is $G$ itself, then line five is triggered. This is checked by comparing \code{s} and \code{v}. If they are equal, then the computation is interrupted by the \code{stop} function and an error message is produced. The error message describes the C-forests which form the problematic hedge structure for the causal effect of the current recursion stage.

\vspace{0.1cm}
\begin{algorithmic}[1]
\setcounter{ALG@line}{5}
	\State \quad \textbf{if}{ $G[\+ S] \in C(G)$,} \textbf{then} 
		\Statex \quad \quad \textbf{return} {$\sum_{v \in \+ s \setminus \+ y} \prod_{V_i \in \+ S}{P(v_i \vert v_\pi^{(i-1)})}$.}
\end{algorithmic}
\vspace{0.1cm}
If the single C-component found on line four is one of the maximal C-components of $G$, then the function \code{id} returns a new distribution object. The \code{sumset} of this object is set to \code{setdiff(s, y)}. The distribution is a product so it must also be set, that \code{recursive = TRUE} for this new object. The objects in the list \code{children} are determined by new recursive calls for every node $V_i$ in $\+ S$. The conditioning nodes are the ones that precede $V_i$ in the topological ordering \code{to}.

\vspace{0.1cm}
\begin{algorithmic}[1]
\setcounter{ALG@line}{6}
	\State \quad \textbf{if}{ $(\exists \+ S^\prime)\+ S \subset \+ S^\prime \text{ such that } G[\+ S^\prime] \in C(G)$,} \textbf{then}
		\Statex \quad \quad \textbf{return} {\textbf{ID}$(\+ y, \+ x \cap \+ s^\prime, \prod_{V_i \in \+ S^\prime}{P(V_i \vert V_\pi^{(i-1)} \cap \+ S^\prime, v_\pi^{(i-1)} \setminus \+ s^\prime), G[\+ S^\prime}])$.}
\end{algorithmic}
If the single C-component found on line four is not one of the maximal C-components of $G$, then it must be a subgraph of some maximal C-component $G[\+ S^\prime]$. Vector \code{s} is replaced by a vector corresponding to the nodes of $\+ S^\prime$, since the nodes of $\+ S$ are no longer required. The function \code{id} is called with the following attributes \code{id(y, intersect(x, s), probability(recursive = TRUE, children = productlist), s.graph, to)}, where \code{s.graph} is the induced subgraph $G[\+ S^\prime]$ and every distribution object in \code{productlist} is constructed by setting \code{var <- s[i]} and \code{cond <- v[0:(ind[i]-1)]} for every node $V_i$ in $\+ S^\prime$.

Algorithm~\ref{alg:identifycond} is also implemented in \pkg{causaleffect} as the function \code{idc}. This function iterates through the nodes \code{z} which it receives as input in addition to the parameters that were previously defined for the \code{id} function. The d-separation condition on line 1 is checked by using the function \code{dSep} from the \pkg{ggm} package. 

\section[Package causaleffect]{Package \pkg{causaleffect}} \label{Sect:packagece}
The primary goal of the \pkg{causaleffect} package is to provide the implementation described in Section~\ref{Sect:implementinR}. The package also provides a means of importing GraphML files into \proglang{R} while retaining any attributes that have been set for the nodes or edges of the graph. 

\subsection[Using causaleffect in R]{Using \pkg{causaleffect} in \proglang{R}}\label{Sect:Rusage}
The primary function which serves as a wrapper for the functions \code{id} and \code{idc} is called \code{causal.effect}. This function can be called as
\begin{Code}
  causal.effect(y, x, z = NULL, G, expr = TRUE)
\end{Code}
where the parameters \code{y}, \code{x} and \code{G} are identical to those of \code{id}. The parameter \code{z} is optional and it is used to represent the conditioning variables of \code{idc}. The initial probability object \code{P} which is a parameter of \code{id} does not have to be specified by the user. In essence, \code{causal.effect} starts from an empty distribution object, and gradually builds the final expression if possible. Also, the topological ordering \code{to} of the function \code{id} is automatically generated by the \code{topological.sort} function of the \pkg{igraph} package.  It is verified, that the vectors \code{y}, \code{x} and \code{z} actually contain nodes that are present in \code{G}. If \code{G} is not a DAG then \code{causal.effect} will also terminate. The last parameter \code{expr} is a logical variable. If assigned to \code{TRUE}, \code{causal.effect} will return the expression in {\LaTeX} syntax. Otherwise, the \code{probability} object used internally by \code{id} is returned, which can be manually parsed by the user to gain the desired output. The function \code{get.expression} is also provided to get a string representation of a \code{probability} object. This function currently supports {\LaTeX} syntax only.

First, \pkg{causaleffect} is loaded to demonstrate the usage of the package.
\begin{Sinput}
R> library("causaleffect")
\end{Sinput}
The \code{causal.effect} function can be utilized without first importing a graph file. One can utilize the \pkg{igraph} package to construct graphs within \proglang{R} itself. This is demonstrated by replicating some of the graphs of Section~\ref{Sect:practice}. The graph of Figure~\ref{kuva:grfGm} is created as follows.
\begin{Sinput}
R> library("igraph")
R> fig1 <- graph.formula(W -+ X, W -+ Z, X -+ Z, Z -+ Y, X -+ Y, Y -+ X, 
+    simplify = FALSE)
R> fig1 <- set.edge.attribute(graph = fig1, name = "description",
+    index = c(5,6), value = "U")
R> ce1 <- causal.effect(y = "Y", x = "X", z = NULL, G = fig1, expr = TRUE)
R> ce1
\end{Sinput}
{\small
\begin{Soutput}
[1] "\\left(\\sum_{W,Z}P(W)P(Z|W,X)\\left(\\sum_{X}P(Y|W,X,Z)P(X|W)\\right)\\right)"
\end{Soutput}
}
Here \code{X -+ Z} denotes a directed edge from $X$ to $Z$. The argument \code{simplify = FALSE} allows the insertion of duplicate edges for the purposes of forming bidirected arcs. Recalling the \code{internal} notation from Section~\ref{Sect:files} we must denote the unidirected edges that correspond to a bidirected edge with a special \code{description} parameter, and assign its value to \code{"U"}. This can be done with the \code{set.edge.attribute} function of the \pkg{igraph} package. Finally, the expression for the interventional distribution is obtained by using the \code{causal.effect} function. Usually one needs to apply the standard \proglang{R} function \code{cat} to obtain the expression with only singular slash symbols.
\begin{Sinput}
R> cat(ce1)
\end{Sinput}
\begin{Soutput}
\left(\sum_{W,Z}P(W)P(Z|W,X)\left(\sum_{X}P(Y|W,X,Z)P(X|W)\right)\right)
\end{Soutput}
To observe unidentifiability, the graph of Figure~\ref{kuva:grfF}(\subref{kuva:grfFa}) is also constructed and an attempt is made to identify $P_x(y)$.
\begin{Sinput}
R> fig5 <- graph.formula(Z_1 -+ X, X -+ Z_2, Z_2 -+ Y, Z_1 -+ X, X -+ Z_1,
+    Z_1 -+ Z_2, Z_2 -+ Z_1, Z_1 -+ Y, Y -+ Z_1, X -+ Y, Y -+ X, 
+    simplify = FALSE) 
R> fig5 <- set.edge.attribute(graph = fig5, name = "description", 
+    index = 4:11, value = "U")
R> causal.effect(y = "Y", x = "X", z = NULL, G = fig5, expr = TRUE)
\end{Sinput}
\begin{Soutput}
Error: Graph contains a hedge formed by C-forests of nodes: 
  {Z_1,X,Z_2} and {Z_2}.
\end{Soutput}
The identification fails in this case due to a hedge present in the graph.

Another function provided by \pkg{causaleffect} is \code{parse.graphml} which can be called as
\begin{Code}
  parse.graphml(file, format = c("standard", "internal"), nodes = c(), 
    use.names = TRUE)
\end{Code}
Parameter \code{file} is the path to the GraphML file the user wishes to convert into an \pkg{igraph} graph. Parameter \code{format} should match the notation that is used to denote bidirected edges in the input graph. The vector \code{nodes} can be used to give names to the nodes of the graph if they have not been specified in the file itself or alternatively, to replace them. Finally, \code{use.names} is a logical vector indicating whether the names of the nodes should be read from the file or not.

We provide an example GraphML file in the replication materials to demonstrate the use of the \code{parse.graphml} function. The file \code{g1.graphml} contains the graph of Figure~\ref{kuva:grfGm} in \code{standard} notation. This means that we do not have to provide names for the nodes or set the unidentified edges manually. First, we read the file into \proglang{R}. This produces several warnings which can be ignored because they are related to the visual attributes created by the graphical editor that was used to produce \code{g1.graphml}. These attributes play no role in the identification of $P_x(y)$. We omit these warnings from the code for clarity.
\begin{Sinput}
R> gml1 <- parse.graphml("g1.graphml", format = "standard")
R> ce2 <- causal.effect(y = "Y", x = "X", z = NULL, G = gml1, expr = TRUE)
R> cat(ce2)
\end{Sinput}
\begin{Soutput}
\left(\sum_{W,Z}P(W)P(Z|W,X)\left(\sum_{X}P(Y|W,X,Z)P(X|W)\right)\right)
\end{Soutput}
We see that the result agrees with the one derived from the manually constructed graph.

For conditional causal effects, we simply utilize the parameter \code{z} of the \code{causal.effect} function. For example, we can obtain the formula for $P_x(z|w)$ in the graph of Figure~\ref{kuva:grfGm}.
\begin{Sinput}
R> cond1 <- causal.effect(y = "Z", x = "X", z = "W", G = gml1, expr = TRUE)
R> cat(cond1)
\end{Sinput}
\begin{Soutput}
\frac{P(Z|W,X)}{\left(\sum_{Z}P(Z|W,X)\right)}
\end{Soutput}
In mathematical notation the result reads 
$$ \frac{P(z\vert w,x)}{\sum_{z}[P(z\vert w,x)]}.$$
This is a typical case where the resulting expression is slightly awkward due to the incompleteness of the simplification rules. However, in this case it is easy to see that the expression can be simplified into $P(z|w,x)$.

\subsection{A complex expression}\label{Sect:comxplex}

\begin{figure}[t!]
  \vspace*{-0.25cm}
  \centering
  \includegraphics[width=0.35\textwidth]{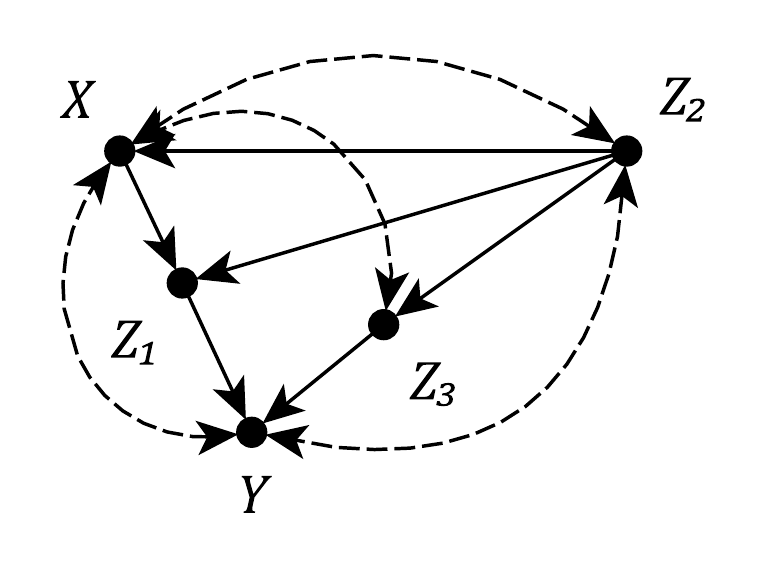}
  \caption{An example of a graph, where an identifiable causal effect results in a complex expression.}
  \label{kuva:grfGtian}
\end{figure}

The conditional distributions $P(v_i \vert v_\pi^{(i-1)})$ that are computed on line 6 can sometimes produce difficult expressions when causal effects are determined from complex graphs. This is a result of the simplification rules which were described in the previous section, and their inability to handle every situation. The graph $G$ of Figure \ref{kuva:grfGtian} serves to demonstrate this phenomenon. An attempt is made to identify $P_x(z_1,z_2,z_3,y)$ in this graph.

\citet{tian02phd} proved this effect to be identifiable, and showed that its expression is
$$P_x(z_1,z_2,z_3,y) = P(z_1|x,z_2)\sum_x P(y,z_3|x,z_1,z_2)P(x,z_2).$$
\noindent
When applying Algorithm~\ref{alg:identify} to this causal effect, it is necessary to compute a conditional distribution $P^*(Y\vert Z_2,Z_3),$ where
$$P^*(y,z_2,z_3) = \sum_{x}P(y\vert z_2,x,z_3,z_1)P(z_3\vert z_2,x)P(x\vert z_2)P(z_2)$$
and $P$ is the joint distribution of the observed variables of $G$.
Now, the function \code{causal.effect} is applied as follows.
\begin{Sinput}
R> fig7 <- graph.formula(X -+ Z_1, Z_1 -+ Y, Z_3 -+ Y, Z_2 -+ X,
+    Z_2 -+ Z_1, Z_2 -+ Z_3, X -+ Y, Y -+ X, X -+ Z_3, Z_3 -+ X, 
+    X -+ Z_2, Z_2 -+ X, Y -+ Z_2, Z_2 -+ Y, simplify = FALSE)
R> fig7 <- set.edge.attribute(graph = fig7, name = "description", 
+    index = 7:14, value = "U")
R> ce3 <- causal.effect(y = c("Z_1", "Z_2", "Z_3", "Y"), x = "X",
+    z = NULL, G = fig7, expr = TRUE)
R> cat(ce3)
\end{Sinput}
This results in the expression
\begin{align*}
 P(z_1|z_2,x)&\frac{\left(\sum_{x}P(y|z_2,x,z_3,z_1)P(z_3|z_2,x)P(x|z_2)P(z_2)\right)}{\left(\sum_{x,y}P(y|z_2,x,z_3,z_1)P(z_3|z_2,x)P(x|z_2)P(z_2)\right)} \\
 &\times \left(\sum_{x,z_3,y}P(y|z_2,x,z_3,z_1)P(z_3|z_2,x)P(x|z_2)P(z_2)\right)P(z_3|z_2)
\end{align*}
This result is clearly more cumbersome than the one determined by Tian. However, it can be shown that this expression is correct by using do-calculus. Because the set $\{X,Z_2\}$ d-separates all paths from $Z_1$ to $Z_3$, it follows that $(Z_3 \indep Z_1|X,Z_2)_G,$ so 
\begin{align*}
 &P(z_1\vert z_2,x)\sum_{x}P(y\vert z_2,x,z_3,z_1)P(z_3\vert z_2,x)P(x\vert z_2)P(z_2) \\ 
 =\;& P(z_1\vert z_2,x)\sum_{x}P(y\vert z_2,x,z_3,z_1)P(z_3\vert z_2,x,z_1)P(x,z_2) \\
 =\;& P(z_1\vert z_2,x)\sum_{x}P(y,z_3\vert z_2,x,z_1)P(x,z_2),
 \end{align*}
where the second equality is due to the conditional independence of $Z_1$ and $Z_3$ given $X$ and $Z_2$. The last line is equivalent with Tian's expression up to the ordering of terms. It can be shown, that the remaining terms are subtracted from the expression.
\begin{align*}
 &\frac{P(z_3\vert z_2)\sum_{x,z_3,y}P(y|z_2,x,z_3,z_1)P(z_3|z_2,x)P(x|z_2)P(z_2)}{\sum_{x,y}P(y\vert z_2,x,z_3,z_1)P(z_3\vert z_2,x)P(x\vert z_2)P(z_2)} \\
 =\;& \frac{P(z_3\vert z_2)P(z_2)}{\sum_{x,y}P(y\vert z_2,x,z_3,z_1)P(z_3\vert z_2,x)P(x, z_2)}.
 \end{align*} 
By applying the same logic to the denominator, it follows that
$$
 \frac{P(z_3\vert z_2)P(z_2)}{\sum_{x,y}P(y\vert z_2,x,z_3,z_1)P(z_3\vert z_2,x)P(x, z_2)} =
\frac{P(z_3\vert z_2)P(z_2)}{\sum_{x,y}P(y,z_3\vert z_2,x,z_1)P(x,z_2)}.
$$
By using the conditional independence of $Z_1$ and $Z_3$ given $X$ and $Z_2$ one gets
 \begin{align*}
 &\frac{P(z_3, z_2)}{\sum_{x}P(z_3\vert z_2,x,z_1)P(x,z_2)} = \frac{P(z_3, z_2)}{\sum_{x}P(z_3\vert z_2,x)P(x,z_2)} \\
 =\;& \frac{P(z_3, z_2)}{\sum_{x}P(z_3\vert z_2,x)P(x,z_2)} 
 = \frac{P(z_3, z_2)}{\sum_{x}P(z_3,z_2,x)} = \frac{P(z_3, z_2)}{P(z_3,z_2)} = 1.
 \end{align*}
The expression produced by \code{causal.effect} is correct despite its complexity.

\subsection{d-separation}\label{Sect:dsep}
\begin{figure}[t!]
  \vspace*{-0.25cm}
  \centering
  \includegraphics[width=0.30\textwidth]{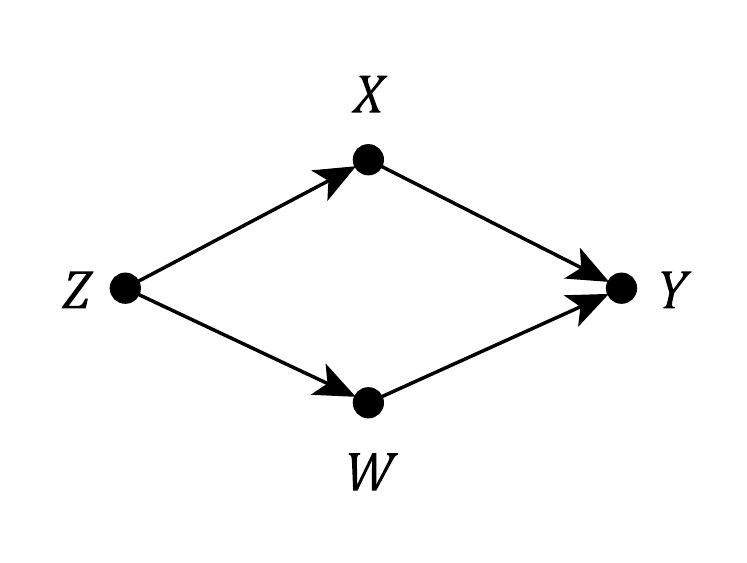}
  \caption{An example of a graph with additional conditional independences.}
  \label{kuva:grfGdsep}
\end{figure}

Algorithm~\ref{alg:identify} does not utilize every possible independence property of a given graph $G$. For example, the conditional distribution of line six is conditioned on all nodes preceding $V_i$ in the topological ordering $\pi$, even though at least some nodes on paths preceding $V_i$ are often d-separated by some sets of nodes. In these cases, the nodes that are d-separated with $V_i$ could be excluded from the expression, because they are conditionally independent from $V_i$ in $G$. This situation is demonstrated by determining the expression of $P_{x,w}(y)$ in the graph $G$ of Figure \ref{kuva:grfGdsep}.

\noindent
The function \code{causal.effect} is utilized
\begin{Sinput}
R> fig8 <- graph.formula(z -+ x, z -+ w, x -+ y, w -+ y)
R> ce3 <- causal.effect(y = "y", x = c("x", "w"), z = NULL, G = fig8,
+    expr = TRUE)
R> cat(ce3)
\end{Sinput}
The function returns $P(y\vert x,w)$ even though Algorithm~\ref{alg:identify} would return $P(y|z,x,w)$. This is possible because $(Y \indep Z \vert X,W)_G$. This means that our implementation is able to simplify the expression into $P(y\vert x,w)$.

\section{Discussion} \label{Sect:discussion}
We have introduced \proglang{R} package \pkg{causaleffect} for deriving expressions of joint interventional distributions in causal models. The task is a specific but important part of causal inference. We believe that our
implementation has two practical use cases. First, \pkg{causaleffect} can be simply used to derive expressions of interventional distributions for complex causal models or to check manual derivations. This is an important step in the estimation of causal effects in complicated settings \citep{karvanen2015}. Second, \pkg{causaleffect} can be used as a building block in simulation studies and automated systems where identifiability needs to be checked for a large number of causal models. An example of this kind usage is already given by \citet{hyttinen2015}.

The efficiency of the presented implementation \pkg{causaleffect} could be analyzed further for example by simulation studies. However, an attempt to maximize performance was made by utilizing the most efficient packages available for the processing of graph files and for the objects corresponding to them. The existing simplification rules of the expressions could also be further improved, but it should be noted that sometimes the more complex expression can prove useful. 

There have been many recent developments in the field of causality resulting in graph theoretic algorithms similar to \textbf{ID} and \textbf{IDC}. These include for example:
\begin{itemize}
\item Causal effect $z$-identifiability algorithm $\text{ID}^\text{Z}$ \citep{Bareinboim:zidentifiability}. $z$-identifiability deals with a situation, where it is possible to utilize a set $\+ Z$ that is disjoint from $\+ X$ to achieve identifiability.
\item Causal effect transportability algorithm sID \citep{bareinboim2013general}. Transportability means, that results obtained from experimental data can be generalized into a larger population, where only observational studies are applicable.
\item Causal effect meta-transportability algorithm $\mu \text{sID}$ \citep{Bareinboim:metatransportability}. Meta-transportability is an extension of the concept of transportability, where the results are to be generalized from multiple experimental studies simultaneously. 
\item Counterfactual and conditional counterfactual identifiability algorithms ID* and IDC* \citep{Shpitser:counterfactuals}.
\end{itemize}
\noindent
The work presented in this paper could be utilized to implement these algorithms.

\bibliography{v76i12}

\newpage

\begin{appendix}

\section{Graphs, causal models and causal effects}\label{App:AppendixA}
\subsection{Graphs}
The definitions that are presented here follow those of \citep{koller09}. \emph{Graph} is an ordered pair $G = \langle \+ V,\+ E \rangle$, where $\+ V$ and $\+ E$ are sets such that 
$$\+ E \subset \{\{X,Y\} \mid X \in \+ V, Y \in \+ V, X \neq Y\}.$$ The elements of $\+ V$ are the nodes of $G$, and the elements of $\+ E$ are the edges of $G$. A graph $F = \langle \+ V^\prime,\+ E^\prime \rangle$ is a \emph{subgraph} of $G$ if $\+ V^\prime \subset \+ V$ and $\+ E^\prime \subset \+ E$. This is denoted as $F \subset G$. A graph $G$ is \emph{directed} if the set $\+ E$ consists of ordered pairs $(X,Y)$. In a directed graph, node $V_2$ is a \emph{child} of node $V_1$ if $G$ contains an edge from $V_1$ to $V_2$, which means that $(V_1,V_2) \in \+ E$. Respectively $V_2$ is a \emph{parent} of $V_1$ if $(V_2,V_1) \in \+ E$. The child-parent relationship is often denoted as $V_1 \rightarrow V_2$, where $V_1$ is a parent of $V_2$ and $V_2$ is a child of $V_1$. This can also be notated as $V_2 \leftarrow V_1$.

\begin{figure}[t!]
  \centering
  \begin{subfigure}[b]{0.3\textwidth}
    \includegraphics[width=1\textwidth]{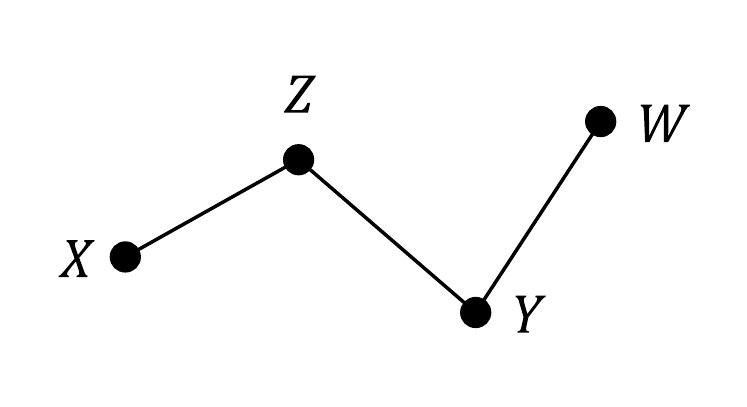}
    \caption{A path.}
    \label{kuva:grfPath}
  \end{subfigure}
  \qquad
  \begin{subfigure}[b]{0.3\textwidth}
    \includegraphics[width=1\textwidth]{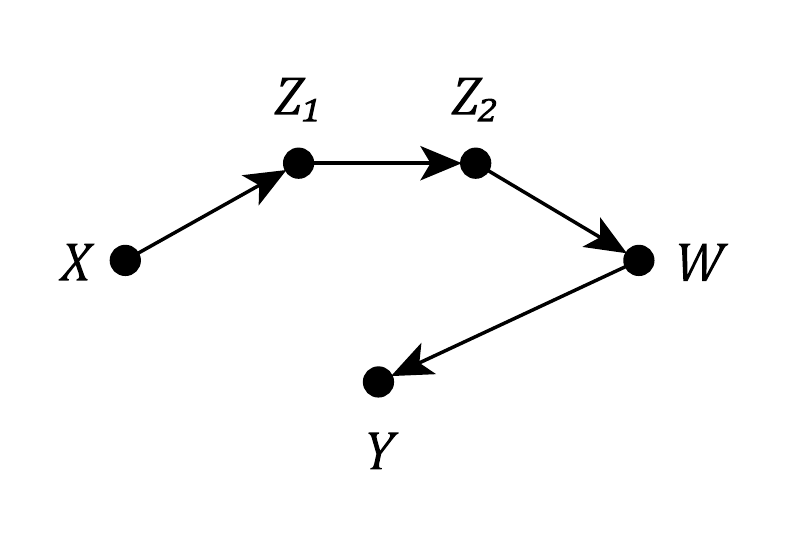}
    \caption{A directed path.}
    \label{kuva:grfDirpath}
  \end{subfigure}
  \\
  \begin{subfigure}[b]{0.3\textwidth}
    \includegraphics[width=1\textwidth]{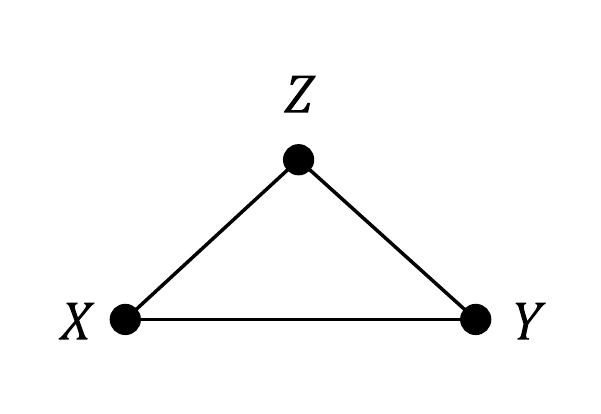}
    \caption{A cycle.}
    \label{kuva:grfCycle}
  \end{subfigure}
  \qquad
  \begin{subfigure}[b]{0.3\textwidth}
    \includegraphics[width=0.95\textwidth]{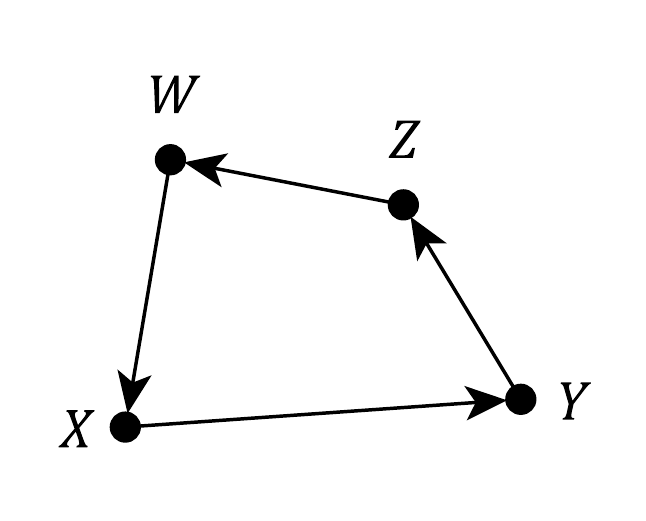}
    \caption{A directed cycle.}
    \label{kuva:grfDircycle}
  \end{subfigure}
  \caption{Directed and undirected paths and cycles.}
  \label{kuva:grfManypaths}
\end{figure}

Let $n \geq 1$, $\+ V = \{V_1,\ldots,V_n\}$ and $V_i \neq V_j$ for all $i \neq j$. If $n > 1$, then the graph $H = \langle \+ V,\+ E \rangle$ is a \emph{path} if 
$$\+ E = \{\{V_1,V_2\},\{V_2,V_3\},\ldots,\{V_{n-1},V_n\}\}$$ 
or if 
$$\+ E = \{\{V_1,V_2\},\{V_2,V_3\},\ldots,\{V_{n-1},V_n\},\{V_n,V_1\}\}.$$ In the first case, $H$ is a path from $V_1$ to $V_n$. In the second case $H$ is a \emph{cycle}. If ${n = 1}$, then $H = \langle \{V_1\},\emptyset \rangle$ is also a path. A path $H$ is a \emph{directed path} if all of its edges are directed and point to the same direction, which means that either
$$\+ E = \{(V_1,V_2),(V_2,V_3),\ldots,(V_{n-1},V_n)\}$$ 
or 
$$\+ E = \{(V_1,V_2),(V_2,V_3),\ldots,(V_{n-1},V_n),(V_n,V_1)\}.$$
A node $V_2$ is a \emph{descendant} of $V_1$ in $G$, if there exists a directed path $H$ from $V_1$ to $V_2$ and $H \subset G$. Respectively, $V_2$ is an \emph{ancestor} of $V_1$ in $G$, if there exists a directed path $H$ from $V_2$ to $V_1$ and $H \subset G$. If a graph $G$ does not contain any cycles, it is \emph{acyclic}. A graph $G = \langle \+ V,\+ E \rangle$ is \emph{connected} if there exists a path $H \subset G$ between every pair of nodes $V_i,V_j \in \+ V$. Examples of paths are cycles are presented in Figure \ref{kuva:grfManypaths}.

\begin{figure}[t!]
	\centering
	\includegraphics[width=0.3\textwidth]{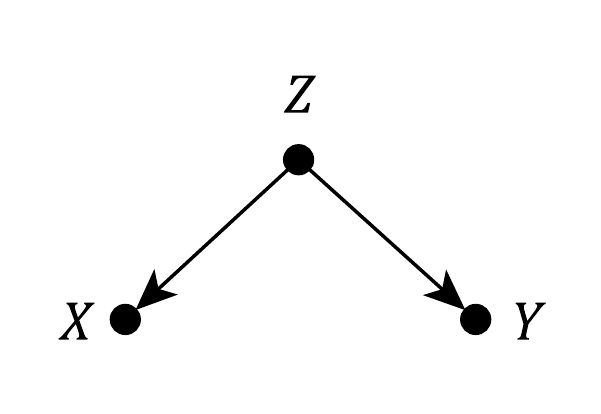}
  \caption{An undirected path in a directed graph.}
  \label{kuva:grfPathSub}
\end{figure}

If a graph is directed it is also possible to consider its subgraphs as undirected graphs, when all of the edges of the graph are regarded as undirected edges. For example, a directed graph contain paths, even if it does not contain any directed paths. The directed graph in Figure \ref{kuva:grfPathSub} contains a path connecting the nodes $X$ and $Y$, even though they are not connected by a directed path.

Let $G = \langle \+ V,\+ E \rangle$ be a graph and $\+ Y \subset \+V$. Assume that the nodes of $\+ Y$ correspond to some observed variables, and that the set $\+ V$ can also contain nodes, which in turn correspond to some unobserved variables. Then the abbreviations $Pa(\+ Y)_G, An(\+ Y)_G, $ and $De(\+ Y)_G$ denote the set of observable parents, ancestors and descendants of the node set $\+ Y$ while also containing $\+ Y$.

\subsection{Causal model}
Causal model can be used to describe the functional relationships between variables of interest. In addition, the model enables the formal treatment of actions or interventions on the variables of the model. Judea Pearl defined the deterministic causal model and its probabilistic counterpart \citep[p.~203-205]{pearl09}, which are presented in this section.

\begin{definition}[Causal Model, \citep{pearl09} 7.1.1]
A \emph{causal model} is a triple
$$ M = \langle \+U,\+V, \+F \rangle,$$
where: 
\begin{enumerate}
\item{$\+U$ is a set of background variables that are determined by factors outside the model;}
\item{$\+V$ is a set $\{V_1,V_2,\ldots,V_n\}$ of variables, called endogenous, that are determined by variables in the model -- that is, variables in $\+U\cup \+ V$; and}
\item{$\+F$ is a set of functions $\{f_{V_1},f_{V_2},\ldots,f_{V_n}\}$ such that each $f_{V_i}$ is a mapping from (the respective domains of) $\+U \cup (\+ V \setminus V_i)$ to $V_i$, and such that the entire set $\+ F$ forms a mapping from $\+ U$ to $\+ V$. In other words, each $f_i$ tells the value of $V_i$ given the values of all other variables in $\+ U \cup \+ V$, and the entire set $\+ F$ has a unique solution $V(u)$. Symbolically, the set of equations $\+ F$ can be represented by writing 
$$ v_i = f_{V_i}(\+ {pa}_{V_i}, \+ u_{V_i}), \quad i = 1,\ldots,n, $$
where $\+{pa}_i$ is any realization of the unique minimal set of variables $\+ {PA}_{V_i}$ in $\+ V \setminus V_i$ (connoting parents) sufficient for representing $f_i$. Likewise, $\+ U_{V_i} \subseteq \+ U$ stand for the unique minimal set of variables in $\+ U$ sufficient for representing $f_i$. }
\end{enumerate}
\end{definition}

For each causal model $M$ there is a corresponding graph $G = \langle \+ W,\+ E \rangle$. The node set $\+ W$ contains a node for each observed and unobserved variable of $M$. The edge set $\+ E$ is determined by the functional relationships between the variables of $\+ V$ and $\+ U$ in the causal model $M$. The set $\+E$ contains an edge from $X$ to $Y$ if $X \in \+{PA}_Y$, which means that there is an edge coming into $V_i$ from every node required to uniquely define $f_{V_i}$. Likewise, the set $\+ E$ contains an edge from $U$ to every node $V_i$ such that $U \in \+ U_{V_i}$.

The definition of causal model does not set any limitations for the unobserved variables. Thus any unobserved node can be a parent of an arbitrary number of observed nodes. If every unobserved node is a parent of exactly two observed nodes, then the causal model is a \emph{semi-Markovian causal model}. \citet{verma93} showed, that for any causal model with unobserved variables one can construct a semi-Markovian causal model that encodes the same set of conditional independences. This is why only semi-Markovian models are considered in this paper. 

\begin{figure}[t!]
	\centering
	\includegraphics[width=0.3\textwidth]{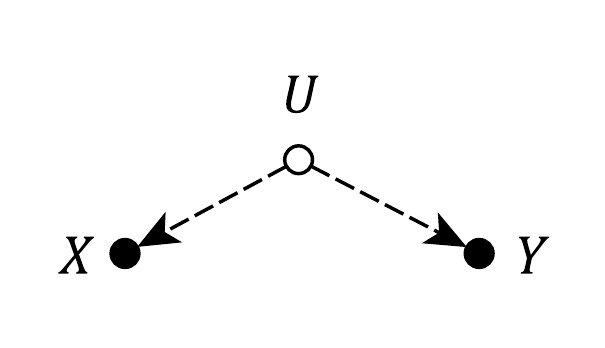}
  \caption{Example notation of unobserved edges.}
  \label{kuva:grfLatent}
\end{figure}

\begin{figure}[t!]
	\centering
	\includegraphics[width=0.3\textwidth]{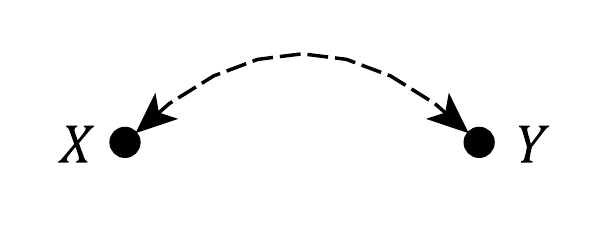}
  \caption{Notation for bidirected edges.}
  \label{kuva:grfShort}
\end{figure}

The edges coming from unobserved variables are sometimes denoted as in Figure \ref{kuva:grfLatent}.
\noindent 
However, it is common not to include the unobserved nodes in the visual representation of the graph, which serves to simplify the notation. Instead, it is said that there exists a \emph{bidirected edge} between $X$ and $Y$, which corresponds to the effect of the unobserved variable. Thus the notation of Figure \ref{kuva:grfShort} is utilized instead of the one in Figure \ref{kuva:grfLatent}.

\noindent 
This notation is used in \citep{Pearlsiscomplete,shpitser06,tian02phd}. It should be noted, that a bidirected edge is not the same as two directed edges between two nodes, as this would induce a cycle in the graph which is not allowed.
Next, the definition of the causal model is expanded by defining a probability distribution for the unobserved variables.

\begin{definition}[Probabilistic Causal Model, \citep{pearl09} 7.1.6]
A \emph{Probabilistic causal model} is a pair 
$$ M = \langle M_D, P(\+ U)\rangle,$$
where $M_D$ is a (deterministic) causal model and $P(\+ U)$ is the joint distribution of the variables in $\+ U$.
\end{definition}

Henceforth in the paper, the term causal model refers to a probabilistic semi-Markovian causal model without exception. Similarly, any graphs discussed will also refer to the graphs induced by these causal models. A graph $G$ induced by a causal model is strongly related to the joint distribution $P$ of all variables in the model, where $P = \prod_{i=1}^nP(v_i\vert pa^*(V_i)_G)\prod_{j=1}^kP(u_j),$ and $Pa^*(.)_G$ also contains all unobserved parents. If this relationship holds, then $G$ is an \emph{I-map} (independence map) of $P$. Independence properties of $G$ and $P$ are closely related through the following definition

\begin{definition}[d-separation, \citep{pearl09} 1.2.3] Let $H = \langle \+ V, \+ E \rangle$ be a path and a set $\+ Z \subset \+ V$. $H$ is said to be \emph{d-separated} by $\+ Z$ in $G$, if and only if either
  \begin{enumerate}
    \item {$H$ contains a chain $I \rightarrow M \rightarrow J$ or a fork $I \leftarrow M \rightarrow J$, where $M \in \+ Z$ and $I,J \in \+ V$., or}
    \item $H$ contains an inverted fork $I \rightarrow M \leftarrow J$, where ${De(M)_G\cap \+ Z = \emptyset}$.
  \end{enumerate}
Disjoint sets $\+ X$ and $\+ Y$ are said to be d-separated by $\+ Z$ in $G$ if every path from $\+ X$ to $\+ Y$ is d-separated by $\+ Z$ in $G$.
\end{definition}
If $\+ X$ and $\+ Y$ are d-separated by $\+ Z$ in $G$, then $\+ X$ is independent of $\+ Y$ given $\+ Z$ in every $P$ for which $G$ is an $I$-map of $P$. The notation of \citep{dawid79} is used to denote this statement as $(\+ X \indep \+ Y\mid \+ Z)_G$.

\subsection{Causal effects} \label{Sect:effect}
Interventions on a causal model alter the functional relationships between its variables. Any intervention $do(\+ X = \+ x)$ on a causal model $M$ produces a new model $M_{\+ x} = \langle \+ U,\+ V,\+ F_{\+ x}, P(\+ U)\rangle$, where $\+ F_{\+ x}$ is obtained by replacing $f_X \in \+ F$ for each $X \in \+ X$ with a constant function, where the constants are defined as the $\+ x$ values of $do(\+ X = \+ x)$. It is now feasible to formalize the notion of causal effects as follows.

\begin{definition}[Causal Effect, \citep{shpitser06}] Let $M = \langle \+ U,\+ V,\+ F, P(\+ U)\rangle$ be a causal model and $\+ Y, \+ X \subset \+ V$. The \emph{causal effect} of ${do(\+ X = \+ x)}$ on the set $\+ Y$ in $M$ is the marginal distribution of $\+ Y$ in $M_{\+ x}$, which is noted by $P(\+ Y\vert do(\+ X = \+ x)) = P_{\+ x}(\+ Y).$
\end{definition}

For every action $do(\+ X = \+ x)$ it is required that $P(\+ x\vert Pa(\+ X)_G\setminus \+ X) > 0$. This limitation ensures that $P_{\+ x}(\+ V)$ and its marginals are well defined. The restriction stems from the fact that it is unfeasible to force $\+ X$ to attain values which cannot be observed. No inference can be made from the distribution of such an intervention using observational data. 

\begin{definition}[Causal Effect Identifiability, \citep{shpitser06} 2] Let $G = \langle \+ V, \+ E \rangle$ be a graph and $\+ Y, \+ X \subset \+ V$. The causal effect of ${do(\+ X = \+ x)}$ on the set $\+ Y$, where $\+ Y \cap \+ X = \emptyset$, is \emph{identifiable} in $G$ if $P_{\+ x}^1(\+ Y) = P_{\+ x}^2(\+ Y)$ for every pair of causal models $M^1$ and $M^2$ such that $P^1(\+ V) = P^2(\+ V)$ and $P^1(\+ x\vert Pa(\+ X)_G\setminus \+ X) > 0$. 
\end{definition}

It is often impossible to show that a causal effect is identifiable by using solely the definition, because one would have to compare every causal model that agree on the distribution of the observed variables. However, the definition serves as a tool to prove unidentifiability in certain cases by constructing two causal models with the same induced graph and observational distribution, and by showing further that the interventional distributions differ. The reader is referred to \citep{shpitser06} for examples.

\end{appendix}

\end{document}